\documentclass[12pt]{article}
\usepackage{amsmath,amssymb,amsthm,amsxtra,overpic,bbm,bm,epsfig,subfigure}
\usepackage{hyperref}
\usepackage{mathrsfs}
\usepackage{enumitem}
\usepackage{graphicx}
\usepackage{color}
\usepackage{comment}
\usepackage{epstopdf}
\usepackage{float}
\usepackage{cite}
\usepackage{slashed,stmaryrd}

\def\thefootnote{\fnsymbol{footnote}}

\usepackage{multirow}
\newcommand{\PL}{P^{}_{\rm L}}
\newcommand{\PR}{P^{}_{\rm R}}
\newcommand{\rmL}{{\rm L}}
\newcommand{\rmR}{{\rm R}}
\newcommand{\rmI}{{\rm i}}
\newcommand{\umass}{\tilde{m}}
\newcommand{\dmass}{\hat{m}}
\newcommand{\amp}{\mathcal{M}}
\newcommand{\lar}{\mathcal{L}}
\newcommand{\cnum}{N^{}_c}
\newcommand{\lam}[2][i]{\lambda^{}_{#1 #2}}
\newcommand{\lamc}[2][i]{\lambda^{\ast}_{#1 #2}}
\newcommand{\lamL}[2][i]{\lambda^{\rm L}_{#1 #2}}
\newcommand{\lamLc}[2][i]{\lambda^{\rm L\ast}_{#1 #2}}
\newcommand{\lamcL}[2][i]{\lambda^{\prime\rm L}_{#1 #2}}
\newcommand{\lamcLc}[2][i]{\lambda^{\prime\rm L\ast}_{#1 #2}}
\newcommand{\lamR}[2][i]{\lambda^{\rm R}_{#1 #2}}
\newcommand{\lamRc}[2][i]{\lambda^{\rm R\ast}_{#1 #2}}
\newcommand{\spinorbar}{\overline{u}\left(p-q\right)}
\newcommand{\spinor}{u\left(p\right)}
\newcommand{\polar}{\epsilon^\ast_\mu \left(q\right)}
\newcommand{\AL}[1][]{\mathcal{A}^{#1}_{\rm L}}
\newcommand{\AR}[1][]{\mathcal{A}^{#1}_{\rm R}}
\newcommand{\asig}{\sigma^{\mu\nu}}

\newcommand{\re}[1]{{\rm Re}\left(#1\right)}
\newcommand{\im}[1]{{\rm Im}\left(#1\right)}

\begin{document}

\vspace{0.2cm}

\begin{center}
{\Large\bf Radiative neutrino masses, lepton flavor mixing
\\and muon $g - 2$ in a leptoquark model}
\end{center}

\vspace{0.2cm}

\begin{center}
{\bf Di Zhang~$^{a,~b}$}~\footnote{E-mail: zhangdi@ihep.ac.cn}
\\
\vspace{0.2cm}
{\small
$^a$Institute of High Energy Physics, Chinese Academy of Sciences, Beijing 100049, China\\
$^b$School of Physical Sciences, University of Chinese Academy of Sciences, Beijing 100049, China}
\end{center}

\vspace{1.5cm}

\begin{abstract}
We propose a leptoquark model with two scalar leptoquarks $S^{}_1  \left( \bar{3},1,\frac{1}{3} \right)$ and $\widetilde{R}^{}_2 \left(3,2,\frac{1}{6} \right)$ to give a combined explanation of neutrino masses, lepton flavor mixing and the anomaly of muon $g-2$, satisfying the constraints from the radiative decays of charged leptons. The neutrino masses are generated via one-loop corrections resulting from a mixing between $S^{}_1$ and $\widetilde{R}^{}_2$. With a set of specific textures for the leptoquark Yukawa coupling matrices, the neutrino mass matrix possesses an approximate $\mu$-$\tau$ reflection symmetry with $\left( M^{}_\nu \right)^{}_{ee} = 0$ only in favor of the normal neutrino mass ordering. We show that this model can successfully explain the anomaly of muon $g-2$ and current experimental neutrino oscillation data under the constraints from the radiative decays of charged leptons.

\end{abstract}


\def\thefootnote{\arabic{footnote}}
\setcounter{footnote}{0}
\newpage
\section{Introduction}\label{sec:intro}
As a quantum field theory with $SU(3)^{}_{\rm c} \times SU(2)^{}_\rmL \times U(1)^{}_{\rm Y}$ gauge symmetry, the Standard Model (SM) has successfully described the behaviors of strong, weak and electromagnetic interactions of all the known fundamental particles. However, there exist some compelling evidences that the SM is incomplete and only an effective field theory at low energy scales at least in its leptonic flavor sector~\cite{Xing:2019vks}. In the neutrino sector, a number of successful neutrino oscillation experiments have firmly proved that neutrinos have tiny masses and the lepton flavor mixing exists~\cite{Zyla:2020zbs}. These facts are in contradiction with the SM and strongly call for new physics beyond the SM. Usually, some new heavy particles are introduced to generate the tiny neutrino masses at the tree level, such as right-handed neutrinos, $SU(2)^{}_{\rm L}$-triplet Higgs and $SU(2)^{}_{\rm L}$-triplet fermions in the type-I~\cite{Minkowski:1977sc,Yanagida:1979as,GellMann:1980vs,Glashow:1979nm,Mohapatra:1979ia}, type-II~\cite{Konetschny:1977bn,Magg:1980ut,Schechter:1980gr,Cheng:1980qt,Lazarides:1980nt,Mohapatra:1980yp} and type-III~\cite{Foot:1988aq,Ma:1998dn} seesaw mechanisms, respectively. An alternative and interesting way to naturally generate the tiny neutrino masses is via radiative corrections~\cite{Zee:1980ai,Zee:1985id,Babu:1988ki} (see. e.g., Refs.~\cite{Cai:2017jrq,Babu:2019mfe}, for a recent review). In the charged-lepton sector, there is a longstanding anomaly of the muon anomalous magnetic moment, namely the discrepancy between the theoretical computations and its measured value, which motivates extensions of the SM with new couplings to leptons. Very recently, the Muon $(g-2)$ Collaboration at Fermi National Laboratory has announced their new result~\cite{Abi:2021gix}, which combined with the previous result from the E821 experiment at Brookhaven National Laboratory~\cite{Bennett:2006fi}, leads to a $4.2\sigma$ discrepancy with the SM prediction~\cite{Aoyama:2020ynm}, that is $\Delta a^{}_\mu = a^{}_\mu \left( \rm Exp \right) - a^{}_\mu \left( \rm SM \right) = (251 \pm 59) \times 10^{-11}$ with $1\sigma$ error
~\footnote{Note that the recent result from lattice QCD is compatible with the experimental value~\cite{Borsanyi:2020mff}.}
. Inspired by this new result, there have been plenty of theoretical works gushing out to explain this discrepancy~\cite{Chiang:2021pma,Lee:2021jdr,Crivellin:2021rbq,Zhang:2021gun,Arcadi:2021cwg,Zhu:2021vlz,Nomura:2021oeu,Endo:2021zal,Das:2021zea,Baum:2021qzx,
Ahmed:2021htr,Ge:2021cjz,Han:2021gfu,Bai:2021bau,Keung:2021rps,Brdar:2021pla,Zu:2021odn,Buen-Abad:2021fwq,
Abdughani:2021pdc,Ferreira:2021gke,Ibe:2021cvf,Babu:2021jnu,VanBeekveld:2021tgn,Chen:2021jok,Cox:2021gqq,
Han:2021ify,Cadeddu:2021dqx,Li:2021poy,Wang:2021fkn,Gu:2021mjd,Wang:2021bcx,Calibbi:2021qto,Yin:2021mls,Cao:2021tuh,
Escribano:2021css,Athron:2021iuf,Chun:2021dwx,Aboubrahim:2021rwz,Bhattacharya:2021ggm,Yang:2021duj,Arcadi:2021yyr,
Lu:2021vcp,Li:2021lnz,Cen:2021iwv,Marzocca:2021azj,Du:2021zkq,Zhou:2021vnf,Ban:2021tos,CarcamoHernandez:2021qhf,
Anchordoqui:2021llp,Baer:2021aax,Altmannshofer:2021hfu,Cacciapaglia:2021gff,Dasgupta:2021dnl,Aboubrahim:2021phn,
Jueid:2021avn,Ma:2021fre,FileviezPerez:2021xfw,Ghorbani:2021yiw,Carpio:2021jhu,Arbuzov:2021lob,Alvarado:2021nxy
,Chang:2021axw,Dutta:2021afo}. A systematic summary of various new-physics scenarios and a great deal of earlier references can be found in Refs~\cite{Athron:2021iuf,Lindner:2016bgg}.

Motivated by above facts, we attempt to extend the SM to give a combined explanation of the tiny neutrino masses, lepton flavor mixing and muon $g-2$. Unfortunately, the most popular way to generate neutrino masses via the three typical seesaw mechanisms can not account for the anomaly of muon $g-2$~\cite{Zhou:2021vnf,Biggio:2008in}. Thus in this work, we focus on an interesting scenario of radiative neutrino masses, where two scalar leptoquarks (LQs), $S^{}_1  \left( \bar{3},1,\frac{1}{3} \right)$ and $\widetilde{R}^{}_2 \left(3,2,\frac{1}{6} \right)$ with the numbers in parenthesis denoting $SU(3)^{}_{\rm c} \times SU(2)^{}_\rmL \times U(1)^{}_{\rm Y}$ quantum numbers are introduced into the SM. LQ extensions of the SM have been discussed extensively due to its possible origination from the grand unification framework~\cite{Pati:1973uk,Pati:1974yy,Fritzsch:1974nn,Georgi:1974sy} and its ability to explain neutrino masses via radiative corrections, the muon $g-2$ and lepton flavor universality violations in semi-leptonic $B$-meson decays~\cite{Lee:2021jdr,Nomura:2021oeu,Keung:2021rps,Athron:2021iuf,Marzocca:2021azj,Du:2021zkq,Ban:2021tos,
FileviezPerez:2021xfw,Chang:2021axw,Djouadi:1989md,Davidson:1993qk,Couture:1995he,Hirsch:1996qy,Hirsch:1996ye,
Chua:1999si,Mahanta:1999xd,Cheung:2001ip,AristizabalSierra:2007nf,Babu:2010vp,Pas:2015hca,Bauer:2015knc,
Cheung:2016fjo,ColuccioLeskow:2016dox,Dorsner:2017wwn,Cai:2017wry,Angelescu:2018tyl,Bigaran:2019bqv,
Dorsner:2019itg,Saad:2020ucl,Dev:2020qet,Saad:2020ihm,Dorsner:2020aaz,Crivellin:2020tsz,Gherardi:2020qhc,
Babu:2020hun,Crivellin:2020mjs} (see, e.g., Ref.~\cite{Dorsner:2016wpm}, for a recent review of LQs). In this work, we only focus on the combined explanation of neutrino masses, lepton flavor mixing and the anomaly of muon $g-2$, as well as constraints from the lepton-flavor-violation (LFV) radiative decays of charged leptons, by means of $S^{}_1$ and $\widetilde{R}_2$. It will show that our model works well and differs from the previous works at least in following three aspects:
\begin{itemize}
  \item To our best knowledge, it is the first time that the combination of $S^{}_1$ and $\widetilde{R}_2$ is used to simultaneously interpret the tiny neutrino masses and the anomaly of muon $g-2$, including the constraints from radiative decays of charged leptons, though the radiative neutrino masses generated by $S^{}_1$ and $\widetilde{R}_2$ and contributions from $S^{}_1$ to the muon $g-2$ have been separately discussed before.
  \item It is well-known that the mixing between $S^{}_1$ and $\widetilde{R}_2$ induced by a LQ-Higgs interaction plays a greatly important role in radiatively generating neutrino masses~\cite{Chua:1999si,Mahanta:1999xd}. This mixing also has significant effects on the branching ratios of radiative charged-lepton decays and then the muon $g-2$. Thus we recalculate all contributions from the physical $S^{}_1$ and $\widetilde{R}_2$ to the branching ratios of radiative charged-lepton decays and the muon $g-2$ in this work, after the mixing between non-physical $S^{}_1$ and $\widetilde{R}_2$ is absorbed into the redefinitions of physical fields.
  \item Motivated by the current experimental data for neutrino oscillations, a set of specific structures for the leptoquark Yukawa coupling matrices are taken to make the neutrino mass matrix approximately possess the $\mu$-$\tau$ reflection symmetry~\cite{Harrison:2002et,Xing:2015fdg}. Given the Yukawa coupling matrices in Eq.~(\ref{eq:Yukawa-matrices}), the neutrino mass matrix additionally has the texture with one zero element, i.e., $\left( M^{}_\nu \right)^{}_{ee} = 0$, which rules out the inverted neutrino mass ordering, and in the limit of $\mu$-$\tau$ reflection symmetry leads to the lightest neutrino mass $m^{}_1$ determined fully by values of $\theta^{}_{12}$ and $\theta^{}_{13}$, besides $\theta^{}_{23}=45^\circ$, $\delta^{}_{\rm CP} = \pm 90^\circ$ and $\rho,\sigma=0$ or $90^\circ$. We analytically discussed the corresponding neutrino masses and lepton flavor mixing induced by this novel neutrino mass matrix with both the exact and approximate $\mu$-$\tau$ reflection symmetry in detail.
\end{itemize}

The rest of this paper is organized as follows. In Sec.~\ref{sec:LQs}, we construct the LQ model with $S^{}_1$ and $\widetilde{R}^{}_2$, and calculate the radiative neutrino mass matrix at the one-loop level, the branching ratios of radiative decays of charged leptons and the muon anomalous magnetic moment. With specific textures of the leptoquark Yukawa coupling matrices, the neutrino masses and lepton flavor mixing with an approximate $\mu$-$\tau$ reflection symmetry are discussed in detail in Sec.~\ref{sec:mass-mixing}. The numerical calculations are carried out in Sec.~\ref{sec:numerical}. Finally, we make a summary in Sec.~\ref{sec:conc}.

\section{The LQ model}\label{sec:LQs}
Enforcing baryon number conservation (e.g., assigning baryon number $B=-\frac{1}{3}$ to $S^{}_1$ and $\widetilde{R}^\ast_2$ under a global $U(1)_{\rm B}$ symmetry) to avoid potentially dangerous proton decay~\footnote{At the renomalizable level, the baryon-number-violating diquark couplings (i.e., $\overline{Q^c_{i \rm L}} \epsilon Q^{}_{j \rm L} S^\dagger_1$ and $\overline{u^c_{i\rm R}} d^{}_{j \rm R} S^\dagger_1$ with $i,j$ being flavor indices)
are forbidden by the global $U(1)^{}_{\rm B}$ symmetry. Note that the dimension five operators involving the leptoquark $\widetilde{R}_2$ (i.e., $\epsilon^{}_{abe} u^a_{i \rm R} d^b_{j \rm R} ( H^\dagger \widetilde{R}^e_2)/\Lambda$ and $\epsilon^{}_{abe} u^a_{i \rm R} e^{}_{\alpha \rm R} (\widetilde{R}^b_2 \epsilon \widetilde{R}^e_2) /\Lambda $ with $i,j,\alpha$ and $a,b,e$ being flavor indices and color indices, respectively) also lead to proton decay, but such dimension five operators can be forbidden by embedding the SM in a larger gauge symmetry group with the right symmetry properties~\cite{Arnold:2013cva,Murgui:2021bdy}.}, the Lagrangian associated with $S^{}_1 \left( \bar{3},1,\frac{1}{3} \right)$ and $\widetilde{R}^{}_2 \left(3,2,\frac{1}{6} \right)$ is given by
\begin{eqnarray}\label{eq:lagrangian}
\lar^{}_{\rm LQ} &=& \lamL{\alpha} \overline{Q^c_{i\rmL}} \epsilon \ell^{}_{\alpha \rmL}S^{}_1 + \lamR{\alpha} \overline{u^c_{i\rmR}} E^{}_{\alpha \rmR} S^{}_1 + \lam{\alpha} \overline{d^{}_{i \rmR}} \widetilde{R}^T_2 \epsilon \ell^{}_{\alpha \rmL} + {\rm h.c.,}
\nonumber
\\
&& + \left( D^{}_\mu S^{}_1 \right)^\dagger \left( D^\mu S^{}_1 \right) + \left( D^{}_\mu \widetilde{R}^{}_2 \right)^\dagger \left( D^\mu \widetilde{R}^{}_2 \right) - V^{}_{\rm LQ} \;,
\end{eqnarray}
where as usual, $Q^{}_\rmL$ and $\ell^{}_\rmL$ are the left-handed quark and lepton doublets of $SU(2)^{}_\rmL$, and $u^{}_\rmR$, $d^{}_\rmR$ and $E^{}_\rmR$ are the right-handed up-type quark, down-type quark and charged lepton singlets of $SU(2)^{}_\rmL$, and the Latin and Greek subscripts running over $(1,2,3)$ and $(e,\mu,\tau)$ are quark and lepton flavor indices, respectively. In Eq.~(\ref{eq:lagrangian}), $\epsilon$ is the $SU(2)^{}_\rmL$ Levi-Civita tensor, the superscript ``$c$" stands for charge conjugation (i.e., $\psi^c = C \overline{\psi}^T$ with $C$ being the charge-conjugation matrix), $D^{}_\mu $ is the gauge covariant derivative. The most general renormalizable potential involving $S^{}_1$ and $\widetilde{R}^{}_2$ is
\begin{eqnarray}\label{eq:potential}
V^{}_{\rm LQ} &=& \mu^2_S S^\dagger_1 S^{}_1 + \mu^2_R \widetilde{R}^\dagger_2 \widetilde{R}^{}_2 + \left( \lambda^{}_{\rm mix} S^\ast_1 \widetilde{R}^\dagger_2 H + {\rm h.c.} \right) + \lambda^{}_{HS} \left( H^\dagger H \right) \left( S^\dagger_1 S^{}_1\right) + \lambda^{(1)}_{HR} \left( H^\dagger H \right) \left( \widetilde{R}^\dagger_2 \widetilde{R}^{}_2 \right)
\nonumber
\\
&& + \lambda^{(3)}_{HR} \left( H^\dagger \tau^I H \right) \left( \widetilde{R}^\dagger_2 \tau^I \widetilde{R}^{}_2 \right) + \lambda^{}_S \left( S^\dagger_1 S^{}_1 \right)^2 + \lambda^{(1)}_{R} \left( \widetilde{R}^\dagger_2 \widetilde{R}^{}_2 \right)^2 + \lambda^{(8)}_R \left( \widetilde{R}^\dagger_2 T^A \widetilde{R}^{}_2 \right)^2
\nonumber
\\
&& + \lambda^{(1)}_{SR} \left( S^\dagger_1 S^{}_1 \right) \left( \widetilde{R}^\dagger_2 \widetilde{R}^{}_2 \right) + \lambda^{(8)}_{SR} \left( S^\dagger_1 T^A S^{}_1 \right) \left( \widetilde{R}^\dagger_2 T^A \widetilde{R}^{}_2 \right) \;,
\end{eqnarray}
in which $H$ is the SM Higgs doublet, and $\tau^I$ (for $I=1,2,3$) and $T^A$ (for $A=1,...,8$) are the Pauli and Gell-Mann matrices, respectively. As we will see later, the LQ-Higgs interaction $\lambda^{}_{\rm mix} S^\ast_1 \widetilde{R}^\dagger_2 H$ provids a mixing between $S^{}_1$ and $\widetilde{R}^{-\frac{1}{3}\ast}_2$ after spontaneous gauge symmetry breaking (SSB), which is required to realize lepton number violation and hence radiatively generate neutrino masses. After the neutral Higgs field obtains its vacuum expectation value, i.e., $\langle H^0 \rangle = v/\sqrt{2}$, the mass matrix for $S^{}_1$ and $\widetilde{R}^{-\frac{1}{3}\ast}_2$ in the basis $\left( S^{}_1, \widetilde{R}^{-\frac{1}{3}\ast}_2 \right) $ is found to be
\begin{eqnarray}\label{eq:LQ-mass-matrix}
M^2_{\rm mix} = \left(\begin{matrix} m^2_S & \displaystyle \frac{v}{\sqrt{2}} \lambda^{}_{\rm mix} \cr \displaystyle \frac{v}{\sqrt{2}} \lambda^\ast_{\rm mix} & m^2_R \end{matrix}\right)
\end{eqnarray}
with
\begin{eqnarray}\label{eq:LQ-mass-mix}
m^2_S = \mu^2_S + \frac{v^2}{2} \lambda^{}_{HS} \;,\quad m^2_R = \mu^2_R + \frac{v^2}{2} \left( \lambda^{(1)}_{HR} + \lambda^{(3)}_{HR} \right) \;.
\end{eqnarray}
For simplicity, we take all parameters in Eqs.~(\ref{eq:LQ-mass-matrix}) and (\ref{eq:LQ-mass-mix}) to be real, thus we can make the following transformations
\begin{eqnarray}\label{eq:LQ-trans}
S^{}_1 \to \cos\theta S^{}_1 + \sin\theta \widetilde{R}^{-\frac{1}{3}\ast}_2 \;,\quad \widetilde{R}^{-\frac{1}{3}\ast}_2 \to -\sin\theta S^{}_1 + \cos\theta \widetilde{R}^{-\frac{1}{3}\ast}_2
\end{eqnarray}
with $\tan 2\theta = \sqrt{2} \lambda^{}_{\rm mix} v/\left( m^2_R - m^2_S \right)$ to work in the basis of the physical LQs which are named in the same way as the non-physical leptoquarks. To avoid misunderstandings, we will make it clear when referring to the physical fields. The masses of the physical LQs are given by
\begin{eqnarray}\label{eq:LQ-mass}
M^2_{1,2} = \frac{1}{2} \left[ m^2_S + m^2_R \pm \sqrt{\left( m^2_S - m^2_R \right)^2 + 2 \lambda^2_{\rm mix} v^2} \right] \;,\quad  M^2_3 = \mu^2_R + \frac{v^2}{2} \left( \lambda^{(1)}_{HR} - \lambda^{(3)}_{HR} \right) \;,
\end{eqnarray}
where $M^{}_1, M^{}_2$ and $ M^{}_3$ are the masses of the physical $S^{}_1$, $\widetilde{R}^{-\frac{1}{3}}_2$ and $\widetilde{R}^{+\frac{2}{3}}_2$, respectively.

Given the transformations in Eq.~(\ref{eq:LQ-trans}), the Yukawa couplings involving the physical LQs after SSB are
\begin{eqnarray}\label{eq:Yukawa}
\lar^{}_{\rm Y} &=& \overline{\nu^{}_\alpha} \left( \lamc{\alpha} \sin\theta \PR - \lamL{\alpha} \cos\theta \PL \right) d^{}_i S^{}_1 + \overline{l^c_\alpha} \left( \lamcL{\alpha} \PL  + \lamR{\alpha} \PR \right) \cos\theta u^{}_i S^{}_1
\nonumber
\\
&& - \overline{\nu^{}_\alpha} \left( \lamc{\alpha} \cos\theta \PR + \lamL{\alpha} \sin\theta \PL \right) d^{}_i \widetilde{R}^{-\frac{1}{3}\ast}_2 + \overline{l^c_{\alpha}} \left( \lamcL{\alpha} \PL + \lamR{\alpha} \PR \right) \sin\theta u^{}_i \widetilde{R}^{-\frac{1}{3}\ast}_2
\nonumber
\\
&& + \lam{\alpha} \overline{d^{}_i} \PL l^{}_{\alpha} \widetilde{R}^{+\frac{2}{3}}_2 + {\rm h.c.},
\end{eqnarray}
with $\lambda^{\prime\rmL} = V^T \lambda^\rmL$, where we work in the down-type quark and charged-lepton mass eigenstate bases (i.e., $d^{}_i$ and $l^{}_\alpha$), and the up-type quark fields have been transformed into their mass eigenstates $u^{}_i$ by the Cabibbo-Kobayashi-Maskawa (CKM) matrix $V$.

\begin{figure}[t]
  \centering
  \includegraphics[width=\textwidth]{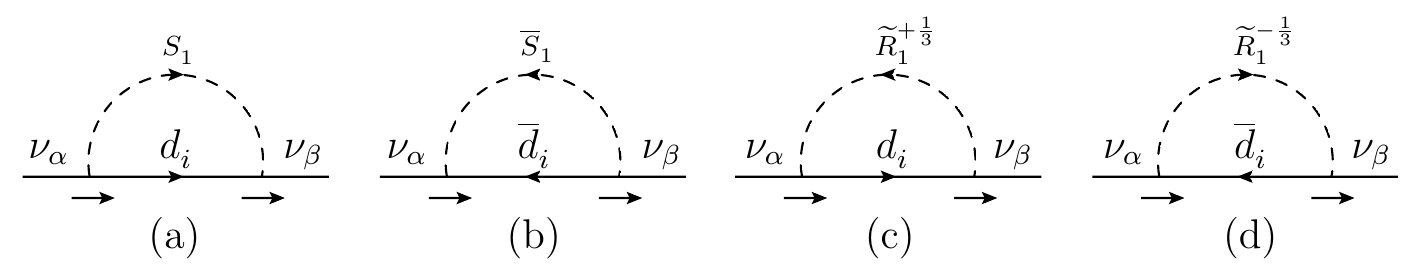}
  \vspace{-0.7cm}
  \caption{Self-energy diagrams for Majorana neutrinos. The arrows below vertices represent the orientations of fermion flow~\cite{Denner:1992me,Denner:1992vza}.}\label{fig:self-energy}
\end{figure}

\subsection{Radiative neutrino masses}\label{sec:radiativemass}
In the considered LQ model described by Eqs.~(\ref{eq:lagrangian}) and (\ref{eq:potential}), neutrinos are exactly massless at the tree level, but they can acquire masses radiatively via the one loop diagrams in Fig.~\ref{fig:self-energy}
~\footnote{All Feynman diagrams in this work are produced by using JaxoDraw~\cite{Binosi:2003yf}.}
. The general structure of the Majorana neutrino self-energy is given by~\cite{Kniehl:1996bd}
\begin{eqnarray}\label{eq:self-energy}
\Sigma^{}_{\alpha\beta} \left( \slashed{p} \right) = \slashed{p} \PL \Sigma^{\rmR\ast}_{\alpha\beta} \left(p^2\right) + \slashed{p} \PR \Sigma^\rmR_{\alpha\beta} \left(p^2\right) + \PL \Sigma^{\rm M\ast}_{\alpha\beta} \left(p^2\right) + \PR \Sigma^{\rm M}_{\alpha \beta} \left(p^2\right) \;,
\end{eqnarray}
where only the last two terms will contribute to neutrino masses. From the Feynman diagrams in Fig.~\ref{fig:self-energy}, one can obtain
\begin{eqnarray}\label{eq:self-energy-amp}
-\rmI \Sigma^{}_{\alpha\beta} \left( \slashed{p} \right) &=& \cnum \int \frac{{\rm d}^4 k}{\left(2\pi\right)^4} \left[ \left( \lamc{\beta} \sin\theta \PR - \lamL{\beta} \cos\theta \PL \right) \frac{1}{\slashed{k}-\dmass^{}_i} \left(\lam{\alpha} \sin\theta \PL - \lamLc{\alpha} \cos\theta \PR \right) \right.
\nonumber
\\
&& \times \frac{1}{\left(k-p\right)^2 - M^2_1} + \left( \lam{\beta} \sin\theta \PL - \lamLc{\beta} \cos\theta \PR \right) \frac{1}{\slashed{k}-\dmass^{}_i} \left(\lamc{\alpha} \sin\theta \PR - \lamL{\alpha} \cos\theta \PL \right)
\nonumber
\\
&& \times \frac{1}{\left(k-p\right)^2 - M^2_1} + \left( \lamc{\beta} \cos\theta \PR - \lamL{\beta} \sin\theta \PL \right) \frac{1}{\slashed{k}-\dmass^{}_i} \left(\lam{\alpha} \cos\theta \PL - \lamLc{\alpha} \sin\theta \PR \right)
\nonumber
\\
&& \times \frac{1}{\left(k-p\right)^2 - M^2_2} +  \left( \lam{\beta} \cos\theta \PL - \lamLc{\beta} \sin\theta \PR \right) \frac{1}{\slashed{k}-\dmass^{}_i} \left(\lamc{\alpha} \cos\theta \PR - \lamL{\alpha} \sin\theta \PL \right)
\nonumber
\\
&& \times \left. \frac{1}{\left(k-p\right)^2 - M^2_2} \right] \;,
\end{eqnarray}
then radiative neutrino mass matrix $M^{}_\nu$ can be extracted from Eq.~(\ref{eq:self-energy-amp}), that is
~\footnote{All one-loop integrals in this work are calculated with the help of Package-X~\cite{Patel:2015tea,Patel:2016fam}.}
\begin{eqnarray}\label{eq:radiative-mass}
\left(M^{}_\nu \right)^{}_{\alpha\beta} &=& \Sigma^{\rm M}_{\alpha\beta} (0) = \frac{\rmI}{2} \cnum \left( \lamc{\alpha} \lamLc{\beta} + \lamLc{\alpha} \lamc{\beta} \right) \sin 2\theta \int \frac{{\rm d}^4 k}{\left(2\pi\right)^4} \frac{\dmass^{}_i}{k^2-\dmass^2_i}
\nonumber
\\
&& \times \left[ \frac{1}{\left(k-p\right)^2-M^2_2} - \frac{1}{\left(k-p\right)^2-M^2_1} \right]
\nonumber
\\
&=& -\frac{\cnum}{2(4\pi)^2} \left( \lamc{\alpha} \lamLc{\beta} + \lamLc{\alpha} \lamc{\beta} \right) \dmass^{}_i \sin2\theta \left[ \frac{M^2_2 \ln \displaystyle \frac{\dmass^2_i}{M^2_2} }{M^2_2 - \dmass^2_i} - \frac{M^2_1 \ln \displaystyle \frac{\dmass^2_i}{M^2_1} }{M^2_1 - \dmass^2_i} \right]
\nonumber
\\
&\simeq&  \frac{3\sin2\theta}{32\pi^2} \ln \frac{M^2_2}{M^2_1} \left[ \left(\lambda^\dagger \right)^{}_{\alpha i} \dmass^{}_i \left( \lambda^{\rmL\ast} \right)^{}_{i\beta} + \left( \lambda^{\rmL\dagger} \right)^{}_{\alpha i} \dmass^{}_i  \left( \lambda^\ast \right)^{}_{i\beta} \right]\;,
\end{eqnarray}
where $\cnum = 3$ is the color number and $\dmass = \left( m^{}_{\rm d}, m^{}_{\rm s}, m^{}_{\rm b} \right)$ denotes the masses of down-type quarks. Note that the mixing between $S^{}_1$ and $\widetilde{R}^{-\frac{1}{3}}_{2}$ is important for radiatively generating neutrino masses, and if the masses of the physical $S^{}_1$ and $\widetilde{R}^{-\frac{1}{3}}_{2}$ are exactly degenerate, three neutrinos will remain massless at the one-loop level.

\begin{figure}[t]
  \centering
  \includegraphics[width=\textwidth]{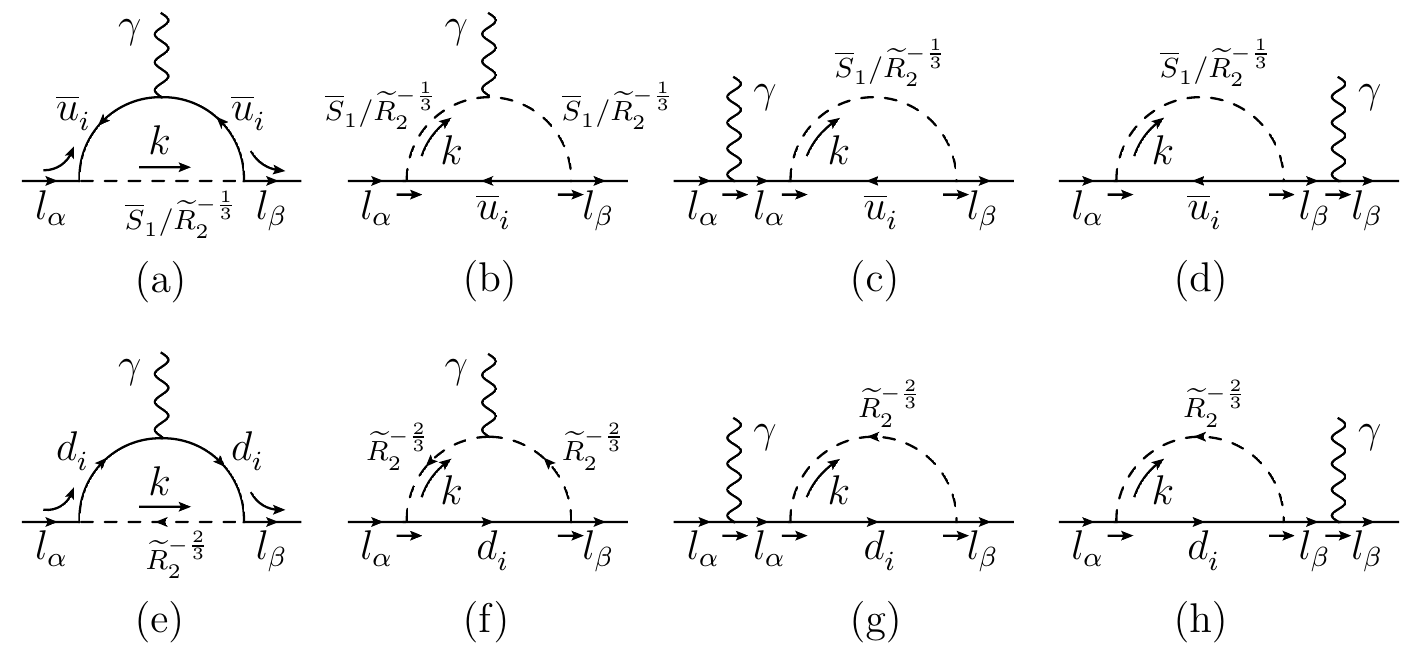}
  \vspace{-0.7cm}
  \caption{Diagrams for the radiative decays of charged leptons, i.e., $l^-_\alpha \to l^-_\beta + \gamma$ with $(\alpha,\beta)= (\tau,\mu)$, $(\tau, e)$ and $(\mu, e)$. The arrows around vertices represent the orientations of fermion flow~\cite{Denner:1992me,Denner:1992vza}.}\label{fig:lfv}
\end{figure}

\subsection{LFV decays and muon $g-2$}\label{sec:AMM}
Induced by the Yukawa couplings of the physical LQs in Eq.~(\ref{eq:Yukawa}), the LFV decays of charged leptons, i.e., $l^-_\alpha \to l^-_\beta + \gamma$ with $(\alpha,\beta)= (\tau,\mu)$, $(\tau, e)$ and $(\mu, e)$ occur at the one-loop level, as shown in Fig.~\ref{fig:lfv}. The muon anomalous magnetic moment can be easily extracted from the total amplitude of the muon radiative decay. Note that all physical LQs will contribute to the total amplitude and lepton anomalous magnetic moment, but the dominant contributions only come from the physical $S^{}_1$ and $\widetilde{R}^{-\frac{1}{3}}_2$ due to the chiral enhancement.

The amplitudes for the radiative decays of charged leptons mediated by the physical $S^{}_1$ as shown by diagrams (a)---(d) in Fig.~\ref{fig:lfv} are given by
\begin{eqnarray}\label{eq:ampa-d}
\rmI \amp^{S^{}_1}_{\rm (a)} &=& \frac{2}{3} \cnum  e \cos^2\theta \polar \spinorbar \left\{ \int \frac{{\rm d}^4 k}{\left(2\pi\right)^4} \left( \lamcLc{\beta} \PR + \lamRc{\beta} \PL \right) \frac{\left( \slashed{p}-\slashed{k} -\slashed{q} + \umass^{}_i \right)}{\left(k+q-p\right)^2 - \umass^2_i} \right.
\nonumber
\\
&& \times \left. \gamma^\mu \frac{ \left( \slashed{p} - \slashed{k} + \umass^{}_i \right)}{ \left(k-p\right)^2 - \umass^2_i}  \left( \lamcL{\alpha} \PL + \lamR{\alpha} \PR \right) \frac{1}{k^2-M^2_1}  \right\}\spinor \;,
\nonumber
\\
\rmI \amp^{S^{}_1}_{\rm (b)} &=& \frac{1}{3} \cnum e \cos^2\theta \polar \spinorbar \left\{ \int \frac{{\rm d}^4 k}{\left(2\pi\right)^4} \left( \lamcLc{\beta} \PR + \lamRc{\beta} \PL \right) \frac{\left( \slashed{p}-\slashed{k}+ \umass^{}_i \right)}{\left(k-p\right)^2 - \umass^2_i} \right.
\nonumber
\\
&& \times \left. \left( \lamcL{\alpha} \PL + \lamR{\alpha} \PR \right) \frac{ 1 }{ \left(k-q\right)^2 - M^2_1} (2k-q)^\mu \frac{1}{k^2-M^2_1}  \right\}\spinor \;,
\nonumber
\\
\rmI \amp^{S^{}_1}_{\rm (c)} &=& -\frac{ \cnum e \cos^2\theta}{m^2_\alpha - m^2_\beta} \polar \spinorbar \left\{ \int \frac{{\rm d}^4 k}{\left(2\pi\right)^4}  \left( \lamcLc{\beta} \PR + \lamRc{\beta} \PL \right) \frac{\left( \slashed{p} -\slashed{k} - \slashed{q} + \umass^{}_i \right)}{\left(k+q-p\right)^2 - \umass^2_i}   \right.
\nonumber
\\
&& \times \left. \left( \lamcL{\alpha} \PL + \lamR{\alpha} \PR \right) \left( \slashed{p} - \slashed{q} + m^{}_\alpha \right) \gamma^\mu \frac{1}{k^2-M^2_1}  \right\}\spinor \;,
\nonumber
\\
\rmI \amp^{S^{}_1}_{\rm (d)} &=&  \frac{ \cnum e \cos^2\theta}{m^2_\alpha - m^2_\beta} \polar \spinorbar \left\{ \int \frac{{\rm d}^4 k}{\left(2\pi\right)^4}  \gamma^\mu \left( \slashed{p} + m^{}_\beta \right) \left( \lamcLc{\beta} \PR + \lamRc{\beta} \PL \right)  \right.
\nonumber
\\
&& \times \left. \frac{\left( \slashed{p}-\slashed{k} + \umass^{}_i \right)}{\left(k-p\right)^2 - \umass^2_i} \left( \lamcL{\alpha} \PL + \lamR{\alpha} \PR \right) \frac{1}{k^2-M^2_1}  \right\}\spinor \;,
\end{eqnarray}
in which $\umass = \left( m^{}_{\rm u}, m^{}_{\rm c}, m^{}_{\rm t} \right)$ and $m = \left( m^{}_e, m^{}_\mu, m^{}_\tau \right)$ denote the masses of up-type quarks and charged leptons, respectively. Then the total amplitude mediated by the physical $S^{}_1$ is found to be
\begin{eqnarray}\label{eq:ampS}
\rmI \amp^{S^{}_1} = -\frac{\rmI e}{\left(4\pi\right)^2} \polar \spinorbar \left( \AL[S^{}_1] \PL + \AR[S^{}_1] \PR \right) \rmI \asig q^{}_\nu \spinor \;,
\end{eqnarray}
where
\begin{eqnarray}\label{eq:ampSALR}
\AL[S^{}_1]  &=& \frac{\cnum \cos^2\theta}{12M^2_1} \left[ 2\umass^{}_i \lamcL{\alpha} \lamRc{\beta} \mathcal{F} \left( \frac{\umass^2_i}{M^2_1} \right) - \left( m^{}_\beta \lamcL{\alpha} \lamcLc{\beta} + m^{}_\alpha \lamR{\alpha} \lamRc{\beta} \right) \mathcal{G} \left( \frac{\umass^2_i}{M^2_1} \right) \right] \;,
\nonumber
\\
\AR[S^{}_1]  &=& \frac{\cnum \cos^2\theta}{12M^2_1} \left[ 2\umass^{}_i \lamR{\alpha} \lamcLc{\beta} \mathcal{F} \left( \frac{\umass^2_i}{M^2_1} \right) - \left( m^{}_\alpha \lamcL{\alpha} \lamcLc{\beta} + m^{}_\beta \lamR{\alpha} \lamRc{\beta} \right) \mathcal{G} \left( \frac{\umass^2_i}{M^2_1} \right) \right] \;,
\end{eqnarray}
with
\begin{eqnarray}\label{eq:loopa-d}
\mathcal{F} (x) = \frac{7 - 8x + x^2 +2(2+x) \ln x }{ (1-x)^3 } \;,\quad
\mathcal{G} (x) = \frac{1 + 4x -5x^2 + 2x(2+x) \ln x }{(1-x)^4} \;.
\end{eqnarray}
If $x \ll 1$ is satisfied, $\mathcal{F}(x)$ and $\mathcal{G}(x)$ may be expanded into $\mathcal{F}(x) \simeq 7 + 4 \ln x$ and $\mathcal{G}(x) \simeq 1$ at the leading order, respectively.

The contributions from the physical $\widetilde{R}^{-\frac{1}{3}}_2$ can be easily obtained by making the replacements $\cos^2\theta \to \sin^2\theta$ and $M^{}_1 \to M^{}_2$ in Eqs.~(\ref{eq:ampS})---(\ref{eq:ampSALR}), that is
\begin{eqnarray}\label{eq:ampR1}
\rmI \amp^{\widetilde{R}^{-\frac{1}{3}}_2} = \rmI \amp^{S^{}_1} (M^{}_1 \to M^{}_2) \tan^2\theta \;,
\end{eqnarray}
with $\AL[\widetilde{R}^{-\frac{1}{3}}_2] = \AL[S^{}_1] (M^{}_1 \to M^{}_2) \tan^2\theta $ and $\AR[\widetilde{R}^{-\frac{1}{3}}_2] = \AR[S^{}_1] (M^{}_1 \to M^{}_2) \tan^2\theta $.

The contributions from $\widetilde{R}^{+\frac{2}{3}}_2$ can be achieved by calculating diagrams (e)---(h) in Fig.~\ref{fig:lfv}, namely
\begin{eqnarray}\label{eq:ampe-h}
\rmI \amp^{\widetilde{R}^{+\frac{2}{3}}_2}_{\rm (e)} &=& \frac{1}{3} \cnum e \lam{\alpha} \lamc{\beta} \polar \spinorbar \left\{ \int \frac{{\rm d}^4 k}{\left( 2\pi \right)^4} \PR \frac{\slashed{p} - \slashed{k} -\slashed{q} + \dmass^{}_i }{ \left( k+q-p \right)^2 - \dmass^2_i } \gamma^\mu \frac{\slashed{p} - \slashed{k} + \dmass^{}_i}{\left(k-p\right)^2 - \dmass^2_i} \PL \frac{1}{ k^2 - M^2_3} \right\} \spinor \;,
\nonumber
\\
\rmI \amp^{\widetilde{R}^{+\frac{2}{3}}_2}_{\rm (f)} &=& \frac{2}{3} \cnum e \lam{\alpha} \lamc{\beta} \polar \spinorbar \left\{ \int \frac{{\rm d}^4 k}{\left( 2\pi \right)^4} \PR \frac{\slashed{p} - \slashed{k} + \dmass^{}_i }{ \left( k-p \right)^2 - \dmass^2_i } \PL \frac{1}{\left(k-q\right)^2 - M^2_3} \frac{1}{ k^2 - M^2_3} \right\} \spinor \;,
\nonumber
\\
\rmI \amp^{\widetilde{R}^{+\frac{2}{3}}_2}_{\rm (g)} &=& - \frac{\cnum e \lam{\alpha} \lamc{\beta}}{m^2_\alpha - m^2_\beta} \polar \spinorbar \left\{ \int \frac{{\rm d}^4 k}{\left( 2\pi \right)^4} \PR \frac{\slashed{p} - \slashed{k} -\slashed{q} + \dmass^{}_i }{ \left( k+q-p \right)^2 - \dmass^2_i } \PL \left( \slashed{p} - \slashed{q} + m^{}_\alpha \right) \gamma^\mu \frac{1}{ k^2 - M^2_3} \right\} \spinor \;,
\nonumber
\\
\rmI \amp^{\widetilde{R}^{+\frac{2}{3}}_2}_{\rm (h)} &=& \frac{\cnum e \lam{\alpha} \lamc{\beta}}{m^2_\alpha - m^2_\beta} \polar \spinorbar \left\{ \int \frac{{\rm d}^4 k}{\left( 2\pi \right)^4} \gamma^\mu \left( \slashed{p} + m^{}_\beta \right) \PR \frac{\slashed{p} - \slashed{k} + \dmass^{}_i }{ \left( k-p \right)^2 - \dmass^2_i } \PL  \frac{1}{ k^2 - M^2_3} \right\} \spinor \;,
\end{eqnarray}
then the total amplitude mediated by $\widetilde{R}^{+\frac{2}{3}}_2$ is given by
\begin{eqnarray}\label{eq:ampR2}
\rmI \amp^{\widetilde{R}^{+\frac{2}{3}}_2} = -\frac{\rmI e}{\left(4\pi\right)^2} \polar \spinorbar \left( \AL[\widetilde{R}^{+\frac{2}{3}}_2] \PL + \AR[\widetilde{R}^{+\frac{2}{3}}_2] \PR \right) \rmI \asig q^{}_\nu \spinor \;,
\end{eqnarray}
where
\begin{eqnarray}\label{eq:ampR2ALR}
\AL[\widetilde{R}^{+\frac{2}{3}}_2] = - \frac{\cnum m^{}_\beta}{12 M^2_3} \lam{\alpha}\lamc{\beta} \mathcal{I} \left( \frac{\dmass^2_i}{M^2_3} \right) \;,\quad
\AR[\widetilde{R}^{+\frac{2}{3}}_2] = - \frac{\cnum m^{}_\alpha}{12 M^2_3} \lam{\alpha}\lamc{\beta} \mathcal{I} \left( \frac{\dmass^2_i}{M^2_3} \right) \;,
\end{eqnarray}
with
\begin{eqnarray}\label{eq:loope-h}
\mathcal{I} (x) = \frac{x\left[ 5 - 4x - x^2 + (2+4x) \ln x \right]}{(1-x)^4} \;,
\end{eqnarray}
and $\mathcal{I} (x) \simeq x \left( 5+2\ln x \right)$ for $x\ll 1$.

Therefore, the total amplitude for the radiative decays of charged leptons is found to be
\begin{eqnarray}\label{eq:amp}
\rmI \amp  = -\frac{\rmI e}{\left(4\pi\right)^2} \polar \spinorbar \left( \AL \PL + \AR \PR \right) \rmI \asig q^{}_\nu \spinor \;,
\end{eqnarray}
with
\begin{eqnarray}\label{eq:ALR}
\AL = \AL[S^{}_1] + \AL[\widetilde{R}^{-\frac{1}{3}}_2] + \AL[\widetilde{R}^{+\frac{2}{3}}_2] \;,\quad
\AR = \AR[S^{}_1] + \AR[\widetilde{R}^{-\frac{1}{3}}_2] + \AR[\widetilde{R}^{+\frac{2}{3}}_2] \;.
\end{eqnarray}
Note that all divergences in the amplitude have been cancelled out, which must be satisfied since there is no corresponding counterterm for charged-lepton radiative decays. With the help of Eq.~(\ref{eq:amp}), one can obtain the branching ratio of $l^-_\alpha \to l^-_\beta + \gamma$,
\begin{eqnarray}\label{eq:branch-ratio}
\mathcal{B} \left( l^-_\alpha \to l^-_\beta + \gamma \right) = \frac{\alpha^{}_{\rm em} \left(m^2_\alpha - m^2_\beta \right)^3}{4\left(4\pi\right)^4 m^3_\alpha \Gamma^{}_\alpha} \left( \left|\AL\right|^2 + \left| \AR \right|^2 \right)
\end{eqnarray}
with $\alpha^{}_{\rm em} = e^2/(4\pi)$ and $\Gamma^{}_\alpha$ being the fine-structure constant and the total decay width of the charged lepton $l^-_\alpha$, respectively. Then the contributions from the LQs to the muon $g-2$ can be easily extracted from Eq.~(\ref{eq:amp}), i.e.,
\begin{eqnarray}\label{eq:AMM}
\Delta a^{}_\mu = \Delta a^{S^{}_1}_\mu + \Delta a^{\widetilde{R}^{-\frac{1}{3}}_2}_\mu + \Delta a^{\widetilde{R}^{+\frac{2}{3}}_2}_\mu \;,
\end{eqnarray}
with
\begin{eqnarray}\label{eq:AMMp}
\Delta a^{S^{}_1}_\mu &=& - \frac{\cnum m^{}_\mu \cos^2\theta}{6\left(4\pi\right)^2 M^2_1} \left[ 2\umass^{}_i \re{ \lamcL{\mu} \lamRc{\mu} } \mathcal{F} \left( \frac{\umass^2_i}{M^2_1} \right) - m^{}_\mu \left( \left|\lamcL{\mu} \right|^2 + \left| \lamR{\mu} \right|^2 \right) \mathcal{G} \left( \frac{\umass^2_i}{M^2_1} \right) \right] \;,
\nonumber
\\
\Delta a^{\widetilde{R}^{-\frac{1}{3}}_2}_\mu &=& - \frac{\cnum m^{}_\mu \sin^2\theta}{6\left(4\pi\right)^2 M^2_2} \left[ 2\umass^{}_i \re{ \lamcL{\mu} \lamRc{\mu} } \mathcal{F} \left( \frac{\umass^2_i}{M^2_2} \right) - m^{}_\mu \left( \left|\lamcL{\mu} \right|^2 + \left| \lamR{\mu} \right|^2 \right) \mathcal{G} \left( \frac{\umass^2_i}{M^2_2} \right) \right] \;,
\nonumber
\\
\Delta a^{\widetilde{R}^{+\frac{2}{3}}_2}_\mu &=& \frac{\cnum m^2_\mu \left| \lam{\mu} \right|^2 }{6\left(4\pi\right)^2 M^2_3} \mathcal{I} \left( \frac{\dmass^2_i}{M^2_3} \right) \;.
\end{eqnarray}

Considering $\umass^{}_i, \dmass^{}_j \ll M^{}_k$ for $i,j,k=1,2,3$ and $m^{}_\beta \ll m^{}_\alpha$ for $(\alpha,\beta) = (\tau, \mu)$, $(\tau, e)$, $(\mu, e)$, the branching ratio in Eq.~(\ref{eq:branch-ratio}) and the muon $g-2$ in Eq.~(\ref{eq:AMM}) are approximately given by
\begin{eqnarray}\label{eq:branch-ratio-approx}
\mathcal{B} \left( l^-_\alpha \to l^-_\beta + \gamma \right) &\simeq& \frac{\alpha^{}_{\rm em} m^3_\alpha}{16\left(4\pi\right)^4 \Gamma^{}_\alpha} \left\{ \left| \umass^{}_i \lamcL{\alpha} \lamRc{\beta} \left[ \frac{\cos^2\theta}{M^2_1} \left(7+4\ln \frac{\umass^2_i}{M^2_1} \right) + \frac{\sin^2\theta}{M^2_2} \left(7+4\ln \frac{\umass^2_i}{M^2_2} \right) \right] \right|^2 \right.
\nonumber
\\
&& + \left. \left| \umass^{}_i \lamR{\alpha} \lamcLc{\beta} \left[ \frac{\cos^2\theta}{M^2_1} \left(7+4\ln \frac{\umass^2_i}{M^2_1} \right) + \frac{\sin^2\theta}{M^2_2} \left(7+4\ln \frac{\umass^2_i}{M^2_2} \right) \right] \right|^2  \right\} \;,
\end{eqnarray}
and
\begin{eqnarray}\label{eq:AMM-approx}
\Delta a^{}_\mu &\simeq& \frac{4m^{}_\mu \umass^{}_i}{\left(4\pi\right)^2} \re{  \lamcL{\mu} \lamRc{\mu}} \left[ \frac{\cos^2\theta}{M^2_1} \left( \ln \frac{M^2_1}{\umass^2_i} - \frac{7}{4} \right) + \frac{\sin^2\theta}{M^2_2} \left( \ln \frac{M^2_2}{\umass^2_i} - \frac{7}{4} \right) \right] \;,
\end{eqnarray}
respectively. It is obvious that both the branching ratio of $l^-_\alpha \to l^-_\beta + \gamma$ and the muon $g-2$ are enhanced by the masses of up-type quarks, especially that of top quark, and if there is no mixing between $S^{}_1$ and $\widetilde{R}^{-\frac{1}{3}}_2$ (i.e., $\theta =0$) or the masses of the physical $S^{}_1$ and $\widetilde{R}^{-\frac{1}{3}}_2$ are exactly degenerate (i.e., $M^{}_1=M^{}_2$), the results in Eqs.~(\ref{eq:branch-ratio-approx}) and (\ref{eq:AMM-approx}) will recover the results caused by $S^{}_1$ all alone~\cite{Cheung:2001ip}.

\section{Neutrino masses and lepton flavor mixing}\label{sec:mass-mixing}
As can be seen in Sec.~\ref{sec:LQs}, the Yukawa coupling matrices $\lambda$ and $\lambda^\rmL$ contribute to the neutrino mass matrix, and both the radiative decays of charged leptons and the muon $g-2$ involve the Yukawa coupling matrices $\lambda^\rmL$ and $\lambda^\rmR$. In order to explain the neutrino oscillation observables and the anomaly of muon $g-2$, meanwhile satisfy the constraints from the radiative decays of charged leptons, we assume that the Yukawa coupling matrices have the following textures:
\begin{eqnarray}\label{eq:Yukawa-matrices}
\lambda = \left(\begin{matrix} 0 & 0 & 0 \cr b (1+\varepsilon) & 0 & c (1+\varepsilon) \cr \displaystyle \frac{m^{}_{\rm s}}{m^{}_{\rm b}} b^\ast & \displaystyle \frac{m^{}_{\rm s}}{m^{}_{\rm b}} c^\ast & d  \end{matrix}\right) \;,\quad \lambda^\rmL = a \left(\begin{matrix} 0 & 0 & 0 \cr 0 & 0 & 1 \cr 0 & 1 & 0 \end{matrix}\right) \;,\quad \lambda^\rmR = a \left(\begin{matrix} 0 & 0 & 0 \cr 0 & x & 0 \cr 0 & 1 & 0 \end{matrix}\right) \;,
\end{eqnarray}
with $\varepsilon$ being a small and real quantity, as well as $a$ and $d$ being real. The textures of $\lambda$ and $\lambda^\rmL$ are motivated by the fact that the lepton flavor mixing approximately has the $\mu$-$\tau$ symmetry~\cite{Xing:2019vks,Xing:2015fdg}. The $(3,2)$ elements of $\lambda^\rmL$ and $\lambda^\rmR$ are responsible for explaining the muon $g-2$, while the $(2,2)$ element of $\lambda^\rmR$ is used to suppress the constraint from the LFV decay of tauon (i.e., $\tau \to \mu +\gamma$), which has been largely enhanced.

By means of Eqs.~(\ref{eq:radiative-mass}) and (\ref{eq:Yukawa-matrices}), one can obtain the neutrino mass matrix, that is
\begin{eqnarray}\label{eq:mass-matrix}
M^{}_\nu = \frac{3 \sin2\theta}{32\pi^2} \ln \left(\frac{M^2_2}{M^2_1} \right) a m^{}_{\rm b} \left(\begin{matrix} 0 & \displaystyle \frac{m^{}_{\rm s}}{m^{}_{\rm b}} b &  \displaystyle \frac{m^{}_{\rm s}}{m^{}_{\rm b}} b^\ast \left( 1+ \varepsilon \right) \vspace{0.2cm}\cr\vspace{0.2cm} \displaystyle \frac{m^{}_{\rm s}}{m^{}_{\rm b}} b & 2\displaystyle \frac{m^{}_{\rm s}}{m^{}_{\rm b}} c & d \cr \displaystyle \frac{m^{}_{\rm s}}{m^{}_{\rm b}} b^\ast \left( 1+ \varepsilon \right)  & d & 2 \displaystyle \frac{m^{}_{\rm s}}{m^{}_{\rm b}} c^\ast (1+\varepsilon)  \end{matrix}\right) \;,
\end{eqnarray}
whose $(1,1)$ element vanishes (i.e., $\left( M^{}_\nu \right)^{}_{ee} = 0$) thanks to the vanishing elements in the first line and the first column of $\lambda^\rmL$ given in Eq.~(\ref{eq:Yukawa-matrices}). It is worth emphasizing  that this novel feature is only in favor of the normal neutrino mass ordering. In the basis where the charged-lepton mass matrix is diagonal, $M^{}_\nu$ can be diagonalized by the Pontecorvo-Maki-Nakagawa-Sakata (PMNS) matrix $U$~\cite{Pontecorvo:1957cp,Maki:1962mu,Pontecorvo:1967fh}, namely, $U^\dagger M^{}_\nu U^\ast = D^{}_\nu$ with $ D^{}_\nu \equiv {\rm Diag} \{ m^{}_1, m^{}_2, m^{}_3\}$. In the standard parametrization~\cite{Zyla:2020zbs}, the PMNS matrix $U$ can be decomposed as
\begin{eqnarray}\label{eq:PMNS}
U = P^{}_l \left(\begin{matrix} c^{}_{12} c^{}_{13} & s^{}_{12}c^{}_{13} & s^{}_{13} e^{-\rmI \delta^{}_{\rm CP}} \cr -s^{}_{12}c^{}_{23} - c^{}_{12}s^{}_{13} s^{}_{23} e^{\rmI \delta^{}_{\rm CP}} & c^{}_{12} c^{}_{23} - s^{}_{12} s^{}_{13} s^{}_{23} e^{\rmI \delta^{}_{\rm CP}} & c^{}_{13} s^{}_{23} \cr s^{}_{12} s^{}_{23} -c^{}_{12}s^{}_{13}c^{}_{23} e^{\rmI \delta^{}_{\rm CP}} & -c^{}_{12}s^{}_{23} - s^{}_{12} s^{}_{13} c^{}_{23} e^{\rmI \delta^{}_{\rm CP}} & c^{}_{13} c^{}_{23} \end{matrix}\right) P^{}_\nu \;,
\end{eqnarray}
where $c^{}_{ij}=\cos\theta^{}_{ij}$ and $s^{}_{ij} = \sin\theta^{}_{ij}$ (for $ij=12,13,23$), $P^{}_l = {\rm Diag} \{ e^{\rmI \phi^{}_e}, e^{\rmI \phi^{}_\mu}, e^{\rmI \phi^{}_\tau} \}$ contains three unphysical phases, and $P^{}_\nu = {\rm Diag} \{ e^{\rmI \rho}, e^{\rmI \sigma}, 1 \}$ is the Majorana phase matrix.

If $\varepsilon = 0$, the neutrino mass matrix $M^{}_\nu$ in Eq.~(\ref{eq:mass-matrix}) will preserve the $\mu$-$\tau$ reflection symmetry~\cite{Harrison:2002et,Xing:2015fdg}. For simplicity, we rewrite $M^{}_\nu$ into
\begin{eqnarray}\label{eq:mu-tau}
M^{\mu-\tau}_\nu = \left(\begin{matrix} 0 & B & B^\ast \cr B & C & D \cr B^\ast & D & C^\ast \end{matrix}\right) \;,
\end{eqnarray}
in the limit of the $\mu$-$\tau$ reflection symmetry, where $B=3\sin2\theta /(32\pi^2) \ln (M^2_2/M^2_1) m^{}_{\rm s} ab$, $C=6\sin2\theta /(32\pi^2) \ln (M^2_2/M^2_1) m^{}_{\rm s} ac$ and $D=3\sin2\theta /(32\pi^2) \ln (M^2_2/M^2_1) m^{}_{\rm b} ad$. It is well-known that the $\mu$-$\tau$ reflection structure leads us to~\cite{Nath:2018hjx,Xing:2019edp}
\begin{eqnarray}\label{eq:mu-tau-mixing}
\theta^{}_{23} = \frac{\pi}{4} \;,\quad \delta^{}_{\rm CP} = \pm \frac{\pi}{2} \;,\quad \rho,\sigma = 0~{\rm or}~\frac{\pi}{2} \;,\quad \phi^{}_e = 0~{\rm or}~\frac{\pi}{2} \;,\quad \phi^{}_\mu + \phi^{}_\tau = 2\phi^{}_e \pm \pi \;.
\end{eqnarray}
In addition, we can obtain simple relationships between the other two mixing angles together with three neutrino masses and the elements of $M^{\mu-\tau}_\nu$, namely~\cite{Xing:2019edp,Aizawa:2004wa,Baba:2010wp,Xing:2010ez}
\begin{eqnarray}\label{eq:relation-1}
\left\{\begin{array}{rcl}
\tan \theta^{}_{13} &=&  \displaystyle \rmI e^{\rmI \delta} \frac{\im{C^\prime}}{\sqrt{2} \re{B^\prime}} \vspace{1mm}
\\
\tan 2\theta^{}_{12} &=& \displaystyle \frac{\sqrt{2} \re{B^\prime} \im{C^\prime}}{c^{}_{13} \left[ \re{C^\prime} \im{C^\prime} - \re{B^\prime} \im{B^\prime} \right]} \vspace{1mm}
\\
m^\rho_1 &=& \displaystyle - D^\prime + \frac{ \re{B^\prime} \im{B^\prime}}{\im{C^\prime}} - \frac{\sqrt{2} \re{B}}{c^{}_{13} \sin{2\theta^{}_{12}}} \vspace{1mm}
\\
m^\sigma_2 &=& \displaystyle - D^\prime + \frac{ \re{B^\prime} \im{B^\prime}}{\im{C^\prime}} + \frac{\sqrt{2} \re{B}}{c^{}_{13} \sin{2\theta^{}_{12}}} \vspace{1mm}
\\
m^{}_3 &=& \displaystyle \frac{2\re{B^\prime} \im{B^\prime}}{\im{C^\prime}}
\end{array}\right.
\Longleftrightarrow
\left\{\begin{array}{rcl}
\re{B^\prime} &=& \displaystyle \frac{ (m^\sigma_2  - m^\rho_1 )c^{}_{13} \sin2\theta^{}_{12} }{2\sqrt{2}} \vspace{1mm}
\\
\im{B^\prime} &=& \displaystyle \rmI e^{\rmI \delta} \frac{m^{}_3 \tan\theta^{}_{13}}{\sqrt{2}} \vspace{1mm}
\\
\re{C^\prime} &=& \displaystyle \frac{\left( m^\sigma_2  - m^\rho_1 \right) \cos2\theta^{}_{12} + m^{}_3}{2} \vspace{1mm}
\\
\im{C^\prime} &=& \displaystyle \rmI e^{\rmI \delta} \frac{\left( m^\sigma_2  - m^\rho_1 \right) s^{}_{13} \sin2\theta^{}_{12}}{2} \vspace{1mm}
\\
D^\prime &=& \displaystyle \frac{m^{}_3 - m^\sigma_2 - m^\rho_1}{2}
\end{array}\right. \;,
\end{eqnarray}
as well as
\begin{eqnarray}\label{eq:relation-2}
m^\rho_1 \cos^2\theta^{}_{12} + m^\sigma_2 \sin^2\theta^{}_{12} - m^{}_3 \tan^2\theta^{}_{13} &=& 0 \;,
\nonumber
\\
\re{B^\prime} \im{C^\prime} \left[ \re{C^\prime} + D^\prime \right] - \im{B^\prime} \left[ 2 \left( \re{B^\prime} \right)^2 - \left( \im{C^\prime} \right)^2 \right] &=& 0 \;,
\end{eqnarray}
where $m^\rho_1 = m^{}_1 \exp\left(2\rmI \rho\right)$, $m^\sigma_2 = m^{}_2 \exp\left(2\rmI \sigma \right)$, $B^\prime = B \exp\left[-\rmI \left( \phi^{}_e + \phi^{}_\mu \right) \right]$, $C^\prime = C \exp\left( -2\rmI \phi^{}_\mu \right)$ and $D^\prime = D \exp \left[ -\rmI \left( \phi^{}_\mu + \phi^{}_\tau \right) \right]$. It is interesting that there exists a relation between the mixing parameters and neutrino masses or between the elements of $M^{\mu-\tau}_\nu$ and unphysical phases, resulting from the absence of $(0,0)$ element of $M^{\mu-\tau}_\nu$, as shown in Eq.~(\ref{eq:relation-2}). Thus the mixing angles and neutrino masses are not independent. If the experimental values of $\Delta m^2_{21}$, $\Delta m^2_{31}$ or $\Delta m^2_{32}$, $\theta^{}_{12}$ and $\theta^{}_{13}$ are taken into account, one will find that only the normal neutrino mass ordering (i.e., $m^{}_1 < m^{}_2 < m^{}_3$) can satisfy the first equation in Eq.~(\ref{eq:relation-2}) with two possible values of $m^{}_1$ corresponding to $\left(\rho,\sigma\right) = (\pi/2,0)$ and $(0,\pi/2)$, respectively, and the inverted neutrino mass ordering (i.e., $m^{}_3 < m^{}_1 < m^{}_2$) is ruled out. By the way, the second equation in Eq.~(\ref{eq:relation-2}) can be used to determinate the unphysical phase $\phi^{}_\mu$ or $\phi^{}_\tau$, only one of which is independent.

Now we consider $\varepsilon$ as a small quantity to break the $\mu$-$\tau$ reflection symmetry, then the neutrino masses and mixing parameters will gain corresponding corrections, namely, $\Delta m^{}_i = m^\prime_i - m^{}_i$ (for $i=1,2,3$), $\Delta \theta^{}_{ij} = \theta^\prime_{ij} - \theta^{}_{ij}$ (for $ij=12,13,23$), $\Delta \delta = \delta^\prime_{\rm CP} - \delta^{}_{\rm CP}$, $\Delta \rho = \rho^\prime -\rho$ and $\Delta \sigma = \sigma^\prime - \sigma$ where the parameters with superscripts ``$\prime$" are the neutrino masses and mixing parameters for $\varepsilon \neq 0$. The PMNS matrix $U^\prime$ to diagonalize the mass matrix with $\varepsilon \neq 0$ in Eq.~(\ref{eq:mass-matrix}) consists of those new mixing parameters and three new unphysical phases $\phi^\prime_\alpha$ (for $\alpha = e,\mu,\tau$). Following the method described in Ref.~\cite{Zhang:2020lsd}, one can obtain
\begin{eqnarray}\label{eq:linear-eqns}
\Delta m^{}_i &\simeq& \re{ U^\dagger  \Delta M^{}_\nu U^\ast }^{}_{ii} \;,
\nonumber
\\
\im{ \Delta U^\dagger U }^{}_{ii} &\simeq& - \frac{\im{ U^\dagger  \Delta M^{}_\nu U^\ast }^{}_{ii} }{ 2 m^{}_i} \;,
\nonumber
\\
\re{ \Delta U^\dagger U }^{}_{jk} &\simeq& \frac{ \re{ U^\dagger  \Delta M^{}_\nu U^\ast }^{}_{jk} }{m^{}_j - m^{}_k} \;,
\nonumber
\\
\im{ \Delta U^\dagger U }^{}_{jk} &\simeq& - \frac{\im{ U^\dagger  \Delta M^{}_\nu U^\ast }^{}_{jk} }{m^{}_j + m^{}_k} \;,
\end{eqnarray}
for $i=1,2,3$ and $jk=12,13,23$, where $\Delta M^{}_\mu \equiv M^{}_\nu - M^{\mu-\tau}_\nu$ and $\Delta U \equiv U^\prime - U$. Solving above linear equations leads us to
\begin{eqnarray}\label{eq:de-mass}
\Delta m^{}_1 &\simeq& \frac{\eta}{2} \varepsilon \left\{ \re{C^\prime} s^2_{12} - \sqrt{2} \re{B^\prime} \sin2\theta^{}_{12} c^{}_{13} + s^{}_{13} \left[ \eta^{}_\delta \im{C^\prime} \sin2\theta^{}_{12} - \re{C^\prime} c^2_{12} s^{}_{13} \right.\right.
\nonumber
\\
&& - \left.\left. 2\sqrt{2} \eta^{}_\delta \im{B^\prime} c^2_{12} c^{}_{13}  \right]  \right\} \;,
\nonumber
\\
\Delta m^{}_2 &\simeq& - \frac{\eta}{2} \varepsilon \left\{ \re{C^\prime} c^2_{12} + \sqrt{2} \re{B^\prime} \sin2\theta^{}_{12} c^{}_{13} - s^{}_{13} \left[ \eta^{}_\delta \im{C^\prime} \sin2\theta^{}_{12} + \re{C^\prime} s^2_{12} s^{}_{13} \right.\right.
\nonumber
\\
&& + \left.\left.  2\sqrt{2} \eta^{}_\delta \im{B^\prime} s^2_{12} c^{}_{13}  \right] \right\} \;,
\nonumber
\\
\Delta m^{}_3 &\simeq& \frac{1}{2} \varepsilon c^{}_{13} \left[ \re{C^\prime} c^{}_{13} - 2\sqrt{2} \eta^{}_\delta \im{B^\prime} s^{}_{13} \right] \;,
\end{eqnarray}
and
\begin{eqnarray}\label{eq:de-mixing}
\Delta \theta^{}_{12} &\simeq& \frac{\varepsilon}{2} \left\{ \frac{1}{2} \sin2\theta^{}_{12} s^{}_{13} \left( \re{C^\prime} s^{}_{13} + \sqrt{2} \eta^{}_\delta \im{B^\prime} \cos2\theta^{}_{13} \sec\theta^{}_{13} \right) \left( \frac{1}{m^{}_3 - \eta m^{}_2 } - \frac{1}{m^{}_3 + \eta m^{}_1} \right)  \right.
\nonumber
\\
&&+ s^{}_{13} \left( \eta^{}_\delta \im{C^\prime} + \sqrt{2} \re{B^\prime} \tan\theta^{}_{13} \right) \left( \frac{s^2_{12}}{m^{}_3 + \eta m^{}_1} + \frac{c^2_{12}}{m^{}_3 - \eta m^{}_2} \right)
\nonumber
\\
&&+ \left[ \frac{1}{2} \sin2\theta^{}_{12} \left( \re{C^\prime} (1+ s^2_{13}) +  \sqrt{2} \eta^{}_\delta \im{B^\prime} \sin2\theta^{}_{13} \right) + \cos2\theta^{}_{12} \right.
\nonumber
\\
&& \times \left.\left. \left( \eta^{}_\delta \im{C^\prime} s^{}_{13} - \sqrt{2} \re{B^\prime} c^{}_{13} \right) \right] \frac{\eta}{m^{}_1+m^{}_2}  \right\} \;,
\nonumber
\\
\Delta \theta^{}_{13} &\simeq& \frac{\varepsilon}{4} \left[ \sin2\theta^{}_{12} \left( \sqrt{2} \re{B^\prime} s^{}_{13} + \eta^{}_\delta \im{C^\prime} c^{}_{13} \right) \left( \frac{1}{m^{}_3 + \eta m^{}_1} - \frac{1}{m^{}_3 - \eta m^{}_2} \right)  \right.
\nonumber
\\
&& - \left. \left( \re{C^\prime} \sin2\theta^{}_{13} + 2\sqrt{2} \eta^{}_\delta \im{B^\prime} \cos2\theta^{}_{13}  \right) \left( \frac{c^2_{12}}{m^{}_3 + \eta m^{}_1} + \frac{s^2_{12}}{m^{}_3 - \eta m^{}_2} \right) \right] \;,
\nonumber
\\
\Delta \theta^{}_{23} &\simeq& \frac{\varepsilon}{2} \left[ \frac{1}{2} \sin2\theta^{}_{12} \left( \sqrt{2} \re{B^\prime} \cos2\theta^{}_{13} \sec\theta^{}_{13} - \eta^{}_\delta \im{C^\prime} s^{}_{13} \right) \left( \frac{1}{m^{}_3 - \eta m^{}_1} - \frac{1}{m^{}_3 + \eta m^{}_2} \right) \right.
\nonumber
\\
&& + \left. \left( \sqrt{2} \eta^{}_\delta \im{B^\prime} \tan\theta^{}_{13} - \re{C^\prime} \right) \left( \frac{c^2_{12}}{m^{}_3 + \eta m^{}_2} + \frac{s^2_{12}}{ m^{}_3 - \eta m^{}_1 } \right) \right] \;,
\nonumber
\\
\Delta \delta^{}_{\rm CP} &\simeq& -\frac{\varepsilon}{2} \left\{ \left( 2\sqrt{2} \eta^{}_\delta \re{B^\prime} \cot2\theta^{}_{13} - \im{C^\prime}  \right) \left( \frac{ \cos2\theta^{}_{12} - c^2_{13} c^2_{12} }{m^{}_3 + \eta m^{}_2} - \frac{\cos2\theta^{}_{12} + c^2_{13}s^2_{12}}{m^{}_3-\eta m^{}_1} \right) \right.
\nonumber
\\
&& + \left( \eta^{}_\delta \re{C^\prime} \csc\theta^{}_{13} - \sqrt{2} \im{B^\prime} \sec\theta^{}_{13} \right) \left[ \frac{ \cot\theta^{}_{12} \left( \cos2\theta^{}_{12} - c^2_{13} c^2_{12} \right) }{ m^{}_3 + \eta m^{}_2} \right.
\nonumber
\\
&& + \left. \frac{\tan\theta^{}_{12} \left(\cos2\theta^{}_{12} + c^2_{13}s^2_{12}\right) }{m^{}_3-\eta m^{}_1} \right] - \left[ 2\cot2\theta^{}_{12} \left( \sqrt{2} \im{B^\prime} c^{}_{13} + \eta^{}_\delta \re{C^\prime} s^{}_{13} \right) \right.
\nonumber
\\
&& + \left.\left. \sqrt{2} \eta^{}_\delta \re{B^\prime} \sin2\theta^{}_{13} - \im{C^\prime} \left( 1+ s^2_{13} \right) \right] \frac{\eta}{m^{}_2 - m^{}_1} \right\} \;,
\nonumber
\\
\Delta \rho &\simeq& \frac{\varepsilon}{4} \left[ 2s^{}_{13} \left( \eta^{}_\delta \re{C^\prime} - \sqrt{2} \im{B^\prime} \tan\theta^{}_{13} \right) \left( \frac{\cos2\theta^{}_{12} \cot\theta^{}_{12}}{m^{}_3 + \eta m^{}_2} + \frac{\sin2\theta^{}_{12}}{m^{}_3-\eta m^{}_1} \right) \right.
\nonumber
\\
&& - 2\left( \sqrt{2} \eta^{}_\delta \re{B^\prime} \cos2\theta^{}_{13} \tan\theta^{}_{13} - \im{C^\prime} s^2_{13} \right) \left( \frac{2c^2_{12}}{m^{}_3 - \eta m^{}_1} - \frac{\cos2\theta^{}_{12}}{m^{}_3 + \eta m^{}_2} \right)
\nonumber
\\
&& - \eta \left( \sqrt{2} \im{B^\prime} c^{}_{13} + \eta^{}_\delta \re{C^\prime} s^{}_{13} \right) \left( \frac{2\cot\theta^{}_{12} \cos2\theta^{}_{12}}{m^{}_2 - m^{}_1} - \frac{\sin2\theta^{}_{12}}{m^{}_1} \right)
\nonumber
\\
&& - \left( \im{C^\prime} c^2_{13} + \sqrt{2} \eta^{}_\delta \re{B^\prime} \sin2\theta^{}_{13} \right) \left( \eta \frac{c^2_{12}}{m^{}_1} + \eta \frac{2c^2_{12}}{m^{}_2-m^{}_1} - \frac{1}{m^{}_3} \right)
\nonumber
\\
&& + \left. \eta \im{C^\prime} \left( \frac{4c^2_{12}}{m^{}_2 - m^{}_1} + \frac{\cos2\theta^{}_{12}}{m^{}_1} \right) \right] \;,
\nonumber
\\
\Delta \sigma &\simeq& \frac{\varepsilon}{4} \left[ 2s^{}_{13} \left( \sqrt{2} \im{B^\prime} \tan\theta^{}_{13} - \eta^{}_\delta \re{C^\prime} \right) \left( \frac{\sin2\theta^{}_{12}}{m^{}_3 + \eta m^{}_2} - \frac{\cos2\theta^{}_{12} \tan\theta^{}_{12}}{m^{}_3 - \eta m^{}_1} \right) \right.
\nonumber
\\
&& - 2\left( \sqrt{2} \eta^{}_\delta \re{B^\prime} \cos2\theta^{}_{13} \tan\theta^{}_{13} - \im{C^\prime} s^2_{13} \right) \left( \frac{\cos2\theta^{}_{12}}{m^{}_3 - \eta m^{}_1} + \frac{2s^2_{12}}{m^{}_3 + \eta m^{}_2} \right)
\nonumber
\\
&& + \eta \left( \sqrt{2} \im{B^\prime} c^{}_{13} + \eta^{}_\delta \re{C^\prime} s^{}_{13} \right) \left( \frac{\sin2\theta^{}_{12}}{m^{}_2} - \frac{2\tan\theta^{}_{12} \cos2\theta^{}_{12}}{m^{}_2 - m^{}_1} \right)
\nonumber
\\
&& + \left( \im{C^\prime} c^2_{13} + \sqrt{2} \eta^{}_\delta \re{B^\prime} \sin2\theta^{}_{13} \right) \left( \frac{1}{m^{}_3} +  \eta \frac{s^2_{12}}{m^{}_2} - \eta  \frac{2s^2_{12}}{m^{}_2-m^{}_1} \right)
\nonumber
\\
&& + \left. \eta \im{C^\prime} \left(\frac{\cos2\theta^{}_{12}}{m^{}_2 } + \frac{4s^2_{12}}{m^{}_2 - m^{}_1}  \right) \right] \;,
\end{eqnarray}
where $\eta^{}_\delta=\pm 1$ for $\delta^{}_{\rm CP} = \pm \pi/2$ and $\eta^{}=\pm 1$ for $\left( \rho, \sigma \right) = \left( 0, \pi/2 \right), \left(\pi/2, 0\right)$. It is worth pointing out that with the help of Eqs.~(\ref{eq:relation-1}) and (\ref{eq:relation-2}), one can achieve all the deviations given in Eqs.~(\ref{eq:de-mass}) and (\ref{eq:de-mixing}) as functions either only of the neutrino masses and mixing angles or only of the elements of $M^{}_\nu$. Here we focus on the deviations of $\theta^{}_{23}$ and $\delta^{}_{\rm CP}$, approximately given by
\begin{eqnarray}\label{eq:deviation-approx}
\Delta \theta^{}_{23} &\simeq& -\frac{\varepsilon}{4} \;,
\nonumber
\\
\Delta \delta^{}_{\rm CP} &\simeq& \frac{\eta^{}_\delta \varepsilon \left[ 1 - r^{}_{12} -\left(1 + r^{}_{12} \right) \cos2\theta^{}_{12} \right] \cot2\theta^{}_{12} }{4\left(1 - r^{}_{12} \right) \theta^{}_{13}} \;,
\end{eqnarray}
at the leading order with $r^{}_{12}=m^{}_1/m^{}_2$, where Eqs.~(\ref{eq:relation-1}) and (\ref{eq:relation-2}) have been taken into account. Note that the deviation of $\theta^{}_{23}$ is simply proportional to $\varepsilon$ and that of $\delta^{}_{\rm CP}$ is enhanced by the smallness of $\theta^{}_{13}$.

\section{Numerical calculations}\label{sec:numerical}
There are thirteen relevant real parameters $\{ M^{}_i, \theta, a, \re{b}, \im{b}, \re{c}, \im{c}, d, \re{x}, \im{x},\\ \varepsilon \}$ (for $i=1,2,3$) involved in our model with the chosen Yukawa coupling matrices in Eq.~(\ref{eq:Yukawa-matrices}) to confront with the muon $g-2$, constraints from LFV decays of charged leptons, neutrino masses and neutrino mixing parameters. For simplicity, we assume that the mixing between $S^{}_1$ and $\widetilde{R}^{-\frac{1}{3}\ast}_2$ is maximal, namely $\theta=\pi/4$, and masses of the physical LQs satisfy $M^{}_1 = M^{}_2/( 1 + r ) = M^{}_3 = 2$ TeV with $r$ defined as $r \equiv (M^{}_2 - M^{}_1) /M^{}_1$. We take $r$ as a small quantity, which means that masses of the physical $S^{}_1$ and $\widetilde{R}^{-\frac{1}{3}\ast}_2$ are nearly degenerate and is required to give the correct scale of neutrino masses. As we will see, the parameter $x$ can be taken to be a complex constant and used to suppress the constraint from $\tau^- \to \mu^- + \gamma$. Thus, only eight parameters $\{ r, a, \re{b}, \im{b}, \re{c}, \im{c}, d, \varepsilon \}$ are left.

The masses of charged fermions used in the numerical calculations are their central values at $M^{}_{Z}$~\cite{Huang:2020hdv}~\footnote{In principle, one should take the pole masses of charged fermions for precision calculations. Here for illustration, we naively make use of their running masses at $M^{}_Z$ to show that our model can well accommodate the low-energy observables we are concerned about. The differences between pole masses and running masses at $M^{}_Z$ which come from high-order corrections in the perturbative calculations have very limited effects on the numerical analyses.}:
\begin{eqnarray}\label{eq:fermion-mass}
m^{}_{\rm u} &=& 1.23~{\rm MeV} \;,\quad m^{}_{\rm c} = 0.620~{\rm GeV} \;,\quad m^{}_{\rm t} = 168.26~{\rm GeV} \;,
\nonumber
\\
m^{}_{\rm d} &=& 2.67~{\rm MeV} \;,\quad m^{}_{\rm s} = 53.16~{\rm MeV} \;,\quad m^{}_{\rm b} = 2.839~{\rm GeV} \;,
\nonumber
\\
m^{}_e &=& 0.48307~{\rm MeV} \;,\quad m^{}_\mu = 0.101766~{\rm MeV} \;,\quad m^{}_\tau = 1.72856~{\rm GeV} \;,
\end{eqnarray}
and the quark mixing parameters under the standard parametrization also take their central values~\cite{Zyla:2020zbs}:
\begin{eqnarray}\label{eq:quark-mixing}
\sin\theta^{\rm q}_{12} = 0.22650 \;,\quad \sin\theta^{\rm q}_{13} = 0.00361 \;,\quad \sin\theta^{\rm q}_{23} = 0.04053 \;,\quad \delta^{\rm q} = 1.196 \;.
\end{eqnarray}
The difference between the SM value and the combined experimental average value of the muon anomalous magnetic moment with $1\sigma$ error is~\cite{Abi:2021gix}
\begin{eqnarray}\label{eq:ex-dam}
\Delta a^{}_\mu = \left( 251 \pm 59 \right) \times 10^{-11} \;,
\end{eqnarray}
and the current experimental upper bounds on the branching ratios of radiative $l^-_\alpha \to l^-_\beta + \gamma$ decays are
\begin{eqnarray}\label{eq:ex-lfv}
\mathcal{B} \left( \mu^- \to e^- + \gamma \right) &<& 4.2 \times 10^{-13} \;,
\nonumber
\\
\mathcal{B} \left( \tau^- \to e^- + \gamma \right) &<& 3.3 \times 10^{-8} \;,
\nonumber
\\
\mathcal{B} \left( \tau^- \to \mu^- + \gamma \right) &<& 4.4 \times 10^{-8} \;,
\end{eqnarray}
at the $90$\% confidence level~\cite{Zyla:2020zbs}. The best-fit values with $1\sigma$ errors of two neutrino mass-squared differences and neutrino mixing parameters in the normal neutrino mass ordering are~\cite{Esteban:2020cvm}
\begin{eqnarray}\label{eq:ex-neu-mixing}
&&\sin^2\theta^{}_{12} = 0.304^{+0.013}_{-0.012} \;,\quad \sin^2\theta^{}_{13} = 0.02221^{+0.00068}_{-0.00062} \;,\quad \sin^2\theta^{}_{23} = 0.570^{+0.018}_{-0.024} \;,\quad \delta^{}_{\rm CP} = \left( 195^{+51}_{-25} \right)^\circ \;,
\nonumber
\\
&& \Delta m^2_{21} = \left(7.42^{+0.21}_{-0.20} \right) \times 10^{-5}~{\rm eV}^2 \;,\quad \Delta m^2_{31} = \left( 2.514^{+0.028}_{-0.027} \right) \times 10^{-3}~{\rm eV}^2 \;.
\end{eqnarray}

To find out the allowed regions of model parameters, with which the predictions for low-energy observables can be compatible with the experimental data shown in Eqs.~(\ref{eq:ex-dam})---(\ref{eq:ex-neu-mixing}), we define the Gaussian-$\chi^2$ function as
\begin{eqnarray}\label{eq:chi2}
\chi^2 = \sum\limits_{j} \left( \frac{q^{}_j \left( p^{}_i \right) - q^{\rm bf}_j}{\sigma^{}_j} \right)^2 \;,
\end{eqnarray}
where $p^{} = \{ r, a, \re{b}, \im{b}, \re{c}, \im{c}, d, \varepsilon \}$ are model parameters, $q= \{ \Delta m^2_{21}, \Delta m^2_{31}, \\\sin^2\theta^{}_{12}, \sin^2\theta^{}_{13}, \sin^2\theta^{}_{23}, \Delta a^{}_\mu \}$ are low-energy observables, $q^{}_j \left(p^{}_i\right)$ represent the model predictions for these observables, $q^{\rm bf}_j$ stand for the best-fit values of these observables given in Eqs.~(\ref{eq:ex-dam}) and (\ref{eq:ex-neu-mixing}), and $\sigma^{}_j$ are the symmetrized $1\sigma$ errors. Here we do not include the information of $\delta^{}_{\rm CP}$ in the $\chi^2$-function as a result of the weak constraint on it from the current global-fit results~\cite{Esteban:2020cvm}. Then minimizing the $\chi^2$-function defined in Eq.~(\ref{eq:chi2}), we can obtain the best-fit values of the model parameters $p^{}_i$, and we shall directly use $N\sigma = \sqrt{\chi^2}$ to  derive the allowed parameter ranges at $N$ standard deviations.

Before carrying out the numerical calculations, we first discuss some features of the muon $g-2$, the radiative decays of charged leptons and the flavor structure of neutrinos with the chosen Yukawa coupling matrices in Eq.~(\ref{eq:Yukawa-matrices}), which will give guidance on the numerical analyses.
\begin{itemize}
  \item Given the Yukawa coupling matrices in Eq.~(\ref{eq:Yukawa-matrices}), only the branching ratio of $\tau^- \to \mu^- + \gamma$ is largely enhanced and has the possibility to exceed the experimental constraint given in Eq.~(\ref{eq:ex-lfv}). To avoid this excess, we take the $(2,2)$ element of $\lambda^\rmR$ to be
      \begin{eqnarray}\label{eq:x}
      x= - \frac{m^{}_{\rm t} V^\ast_{\rm cb} \mathcal{F}\left( \displaystyle\frac{m^2_{\rm t}}{M^2_1}\right) }{ m^{}_{\rm c} V^\ast_{\rm cs} \mathcal{F} \left( \displaystyle \frac{m^2_{\rm c}}{M^2_1} \right) } \simeq -2.59 + 8.2\times10^{-5} \rmI \;,
      \end{eqnarray}
      with the inputs shown in Eqs.~(\ref{eq:fermion-mass}) and (\ref{eq:quark-mixing}). Then by means of Eq.~(\ref{eq:branch-ratio-approx}), the branching ratio of $\tau^- \to \mu^- +\gamma$ is approximately given by
      \begin{eqnarray}\label{eq:bra-approx}
      \mathcal{B} \left( \tau^- \to \mu^- + \gamma \right) \simeq \frac{\alpha^{}_{\rm em} m^3_\tau m^2_{\rm t} \left|V^{}_{\rm cb}\right|^2 a^4 r^2}{16\pi^4 M^4_1 \Gamma^{}_\tau} \left( \frac{\ln \displaystyle\frac{m^2_{\rm t}}{m^2_{\rm c}}}{7 + 4\ln \displaystyle\frac{m^2_{\rm c}}{M^2_1}} \right)^2 \sim  1\times 10^{-6} a^4 r^2 \;,
      \end{eqnarray}
      which can easily satisfy the experimental constraint with small $a$ and $r$.
  \item With the help of Eqs.~(\ref{eq:AMM-approx}), (\ref{eq:Yukawa-matrices}) and (\ref{eq:x}), one may obtain
      \begin{eqnarray}\label{eq:dmm}
      \Delta a^{}_\mu \simeq \frac{m^{}_\mu m^{}_{\rm t} a^2}{4\pi^2 M^2_1} \re{V^{}_{\rm tb} - \frac{V^{}_{\rm cb}  V^{}_{\rm ts}}{V^{}_{\rm cs}}} \left( \ln\frac{M^2_1}{m^2_{\rm t}} - \frac{7}{4} \right) \sim 251 \times 10^{-11} \left(\frac{a}{0.085} \right)^2 \;,
      \end{eqnarray}
      which means that $a \sim 0.085$ can accommodate the experimental value of $\Delta a^{}_\mu$ in Eq.~(\ref{eq:ex-dam}).
  \item The $(1,1)$ element of $M^{}_\nu$ given in Eq.~(\ref{eq:mass-matrix}) is vanishing, i.e., $\left( M^{}_\nu \right)^{}_{ee} =0$, resulting in the elimination of the inverted neutrino mass ordering in this model. This point also can be seen from Eq.~(\ref{eq:relation-2}) when $\varepsilon$ is small. If we substitute the best-fit values of $\Delta m^2_{21}$, $\Delta m^2_{31}$, $\theta^{}_{12}$ and $\theta^{}_{13}$ into Eq.~(\ref{eq:relation-2}) in the limit of $\mu$-$\tau$ reflection symmetry, we may obtain two solutions for $m^{}_1$: $m^{}_1 \simeq 6.3$ meV for $(\rho,\sigma)=(0,\pi/2)$ and $m^{}_1 \simeq 2.2$ meV for $(\rho,\sigma)=(\pi/2,0)$. This infers that there will be two parameter ranges for $m^{}_1$ around these two solutions when $M^{}_\nu$ with a small $\varepsilon$ is considered.
  \item As can be seen in Eq.~{(\ref{eq:mass-matrix})}, the mass scale of neutrinos is largely enhanced by $m^{}_{\rm b}$. Since $a\sim 0.085$ is necessary to explain $\Delta a^{}_\mu$, the masses $M^{}_1$ and $M^{}_2$ need to be highly but not exactly degenerate to get the right mass scale of neutrinos. Making use of the trace of $M^{}_\nu M^\dagger_\nu$, we have
      \begin{eqnarray}\label{eq:r-approx}
      r \sim \frac{16\pi^2}{3 a m^{}_{\rm b}} \sqrt{\frac{1}{2d^2} \left( 3m^2_1 + \Delta m^2_{21} + \Delta m^2_{31} \right)} \sim 8 \times 10^{-7} \;,
      \end{eqnarray}
      for $a\sim 0.085$, $d \sim \mathcal{O}(0.01)$ and $m^{}_1 \sim \mathcal{O} (0.01~{\rm eV})$.
\end{itemize}

\begin{table}[t!]
  \centering
  \begin{tabular}{c|c|ccc}
  \hline\hline
  & & Best-fit value & $1\sigma$ range & $3\sigma$ range
  \\
  \hline
  \multirow{9}{*}{\rotatebox{90}{Model parameter}} & $a/10^{-2}$ & $8.37$ & $ 7.37 \to 9.24$ & $4.60 \to 10.9 $
  \\
  & \multirow{2}{*}{$\re{b}$} & \multirow{2}{*}{$-0.000473$} & $ (-0.00245 \to 0.0000687)$   & $(-0.00260 \to 0.000613) $
  \\
  & & & $\cup (0.0961 \to 0.608)$ &  $\cup(0.0748 \to 0.682)$
  \\
  & $\im{b}$ & $0.0246$ & $-0.00410 \to 0.214$ & $-0.0192 \to 0.241$
  \\
  & $\re{c}$ & $0.000280$ & $-0.0278 \to 0.196$ & $-0.0940 \to 0.237$
  \\
  & $\im{c}$ & $0.0367$ & $0.0243 \to 1.28$ & $0.0146 \to 1.43$
  \\
  & $d$ & $0.00147$ & $0.000990 \to 0.0338$ & $0.000579 \to 0.0366$
  \\
  & $r/10^{-6}$ & $4.10$ & $0.136 \to 5.89$ & $0.108 \to 10.1$
  \\
  & $\varepsilon$ & $-0.276$ & $-0.342 \to -0.143$& $-0.470 \to -0.0249$
  \\
  \hline\hline
  \multirow{15}{*}{\rotatebox{90}{Low-energy observable}} & \multirow{2}{*}{$m^{}_1/{\rm meV}$} & \multirow{2}{*}{$5.73$} & $(2.12 \to 2.58)$ & $(1.70 \to 3.17)$
  \\
   & & & $\cup (5.28 \to 6.17)$ & $\cup (4.44 \to 7.24)$
  \\
   & $\Delta m^2_{21}/10^{-5}~{\rm eV^2}$ & $7.42$ & $7.22 \to 7.61$ & $6.81 \to 8.03$
  \\
   & $\Delta m^2_{31}/10^{-3}~{\rm eV^2}$ & $2.51$ & $2.49 \to 2.54$ & $2.43 \to 2.60$
  \\
   & $\theta^{}_{12}/^\circ$ & $33.46$ & $32.71 \to 34.19$ & $31.13 \to 35.73$
  \\
   & $\theta^{}_{13}/^\circ$ & $8.57$ & $8.45 \to 8.69$ & $8.19 \to 8.94$
  \\
   & $\theta^{}_{23}/^\circ$ & $49.03$ & $47.89 \to 50.17$ & $45.44 \to 52.65$
  \\
   & \multirow{2}{*}{$\delta^{}_{\rm CP}/^\circ$} & \multirow{2}{*}{$242.99$} & $(236.52 \to 249.95)$ & $(224.14 \to 266.88)$
  \\
   & & & $\cup (275.00 \to 279.01)$ & $\cup (270.82 \to 283.67)$
  \\
   & \multirow{2}{*}{$\rho/^\circ$} & \multirow{2}{*}{$8.58$} & $(6.47 \to 10.43)$ & $(1.03 \to 13.50)$
  \\
   & & & $\cup (94.02 \to 96.92)$ & $\cup (90.66 \to 99.91)$
  \\
   & \multirow{2}{*}{$\sigma/^\circ$} & \multirow{2}{*}{$92.27$} & $(0.29 \to 0.55)$ & $(0.05 \to 0.91)$
  \\
   & & & $\cup (91.73 \to 92.72)$ & $\cup (90.28 \to 93.55)$
  \\
   & $\Delta a^{}_\mu/10^{-11}$ & $251.10$ & $194.74 \to 305.97$ & $75.94 \to 426.33$
  \\
   & $\mathcal{B} \left( \tau^- \to \mu^- + \gamma \right) $ & $1.17\times 10^{-18}$ & $(0.71 \to 1.75) \times 10^{-18}$ & $(0.11 \to 3.39)\times 10^{-18}$
   \\
  \hline\hline
  \end{tabular}
  \caption{The best-fit values (with $\chi^2_{\rm min} = 6.7 \times 10^{-6}$) along with the $1\sigma$ and $3\sigma$ ranges of the model parameters and the predicted low-energy observables. The symbol ``$\cup$" stands for the union of two sets.}\label{tab:parameter-space}
\end{table}

\begin{figure}[t]
	\centering
	\includegraphics[width=\textwidth]{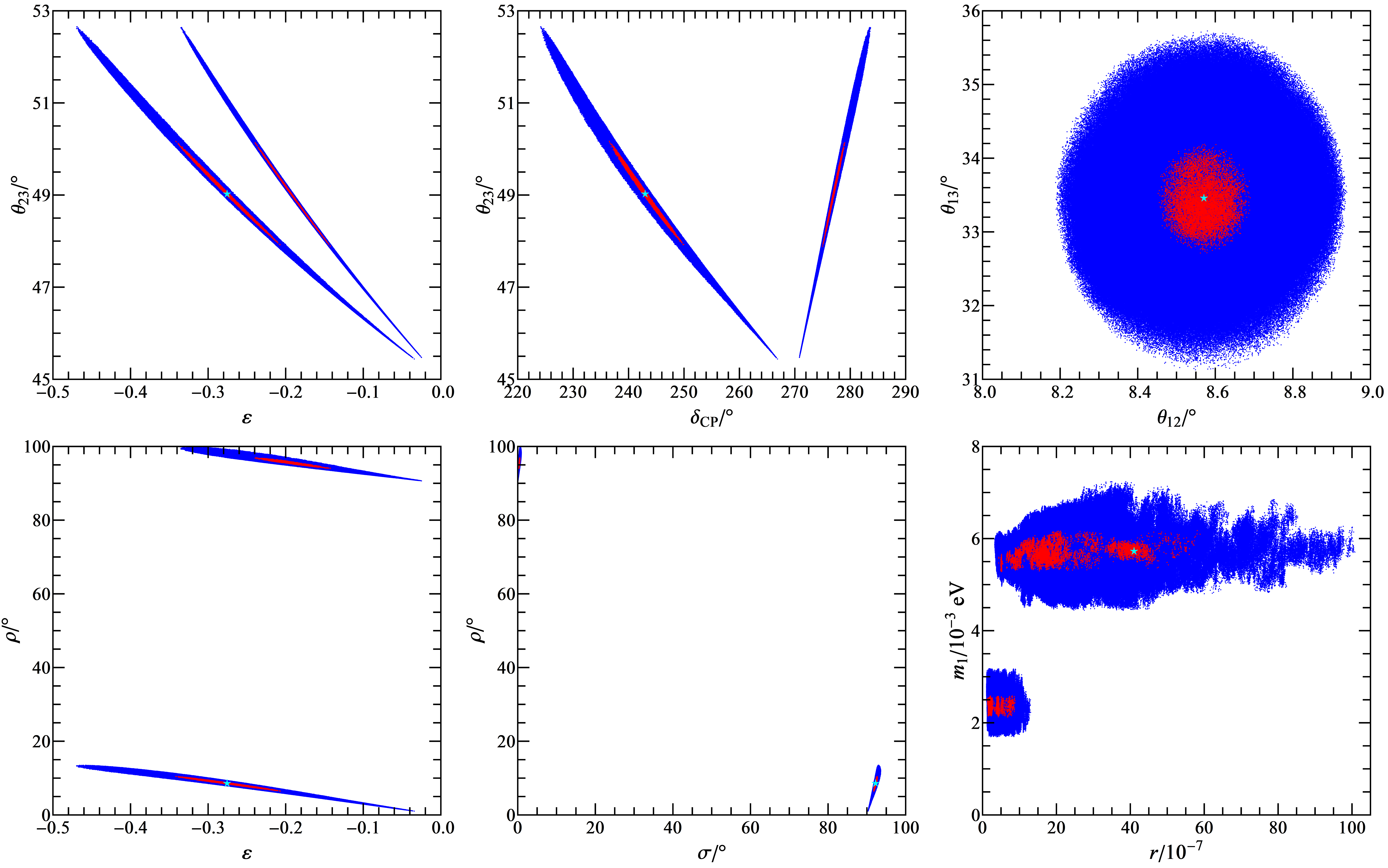}
	\vspace{-0.7cm}
    \caption{Correlations of the pairs $(\theta^{}_{23}, \varepsilon)$, $(\theta^{}_{23}, \delta^{}_{\rm CP})$, $(\theta^{}_{13}, \theta^{}_{12})$, $(\rho, \varepsilon)$, $(\rho, \sigma)$ and $(m^{}_1, r)$, where the cyan stars denote the best-fit values, and the red and blue regions correspond to parameter spaces at $1\sigma$ and $3\sigma$ levels, respectively.}\label{fig:para}
\end{figure}

\begin{figure}[t]
	\centering
	\includegraphics[width=\textwidth]{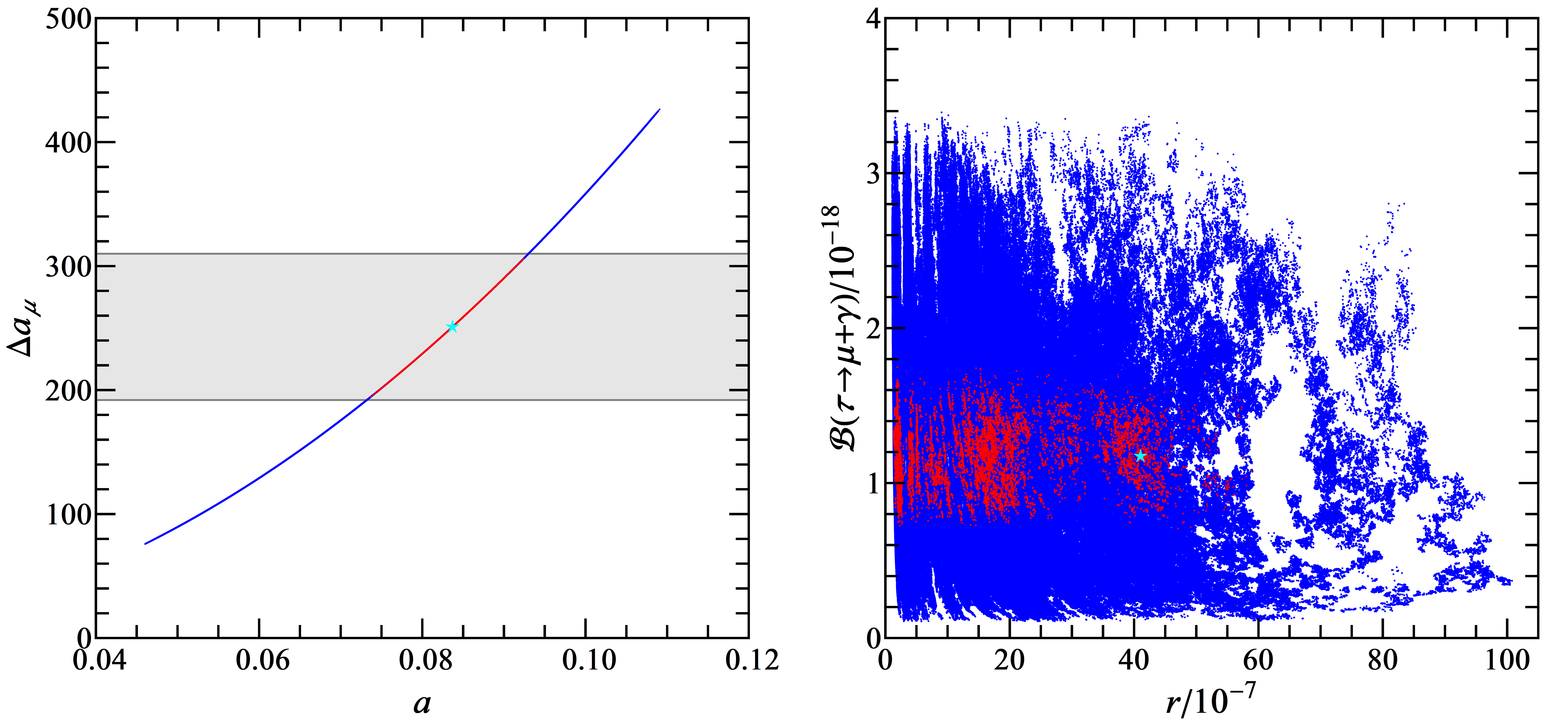}
	\vspace{-0.7cm}
    \caption{Correlation between $\Delta a^{}_\mu$ and $a$, and that between $\mathcal{B} \left( \tau^- \to \mu^- + \gamma \right)$ and $r$, where the cyan stars denote the best-fit values, the gray region in the left panel is the experimental $1\sigma$ range of $\Delta a^{}_\mu$, and the red and blue regions correspond to parameter spaces at $1\sigma$ and $3\sigma$ levels, respectively.}\label{fig:para1}
\end{figure}

Now carrying out the numerical calculations with all inputs, we find that the minimum of $\chi^2$-function is $\chi^2_{\rm min} = 6.7 \times 10^{-6}$ and the corresponding best-fit values of the model parameters and the predicted low-energy observables are listed in the third column of Table~\ref{tab:parameter-space}. In addition, the $1\sigma$ and $3\sigma$ ranges of both the model parameters and the predicted low-energy observables are listed in the last two columns of Table~\ref{tab:parameter-space}, respectively. To make results clearer, we plot correlations of some pairs of the predicted low-energy observables and also those between the predicted low-energy observables and model parameters in Figs.~\ref{fig:para} and \ref{fig:para1}. Some comments on the numerical results are in order.
\begin{itemize}
  \item With the best-fit values of the model parameters listed in the third column of Table~\ref{tab:parameter-space}, our model can successfully reproduce the experimental values of $\Delta m^2_{21}$, $\Delta m^2_{31}$, $\sin^2\theta^{}_{12}$, $\sin^2\theta^{}_{13}$ and $\sin^2\theta^{}_{23}$, and explain the muon anomalous magnetic moment $ \Delta a^{}_\mu$, as well as satisfy the constraints from the LFVs decays of charged leptons, with a quite small value of $\chi^2$-function, namely $\chi^2_{\rm min} = 6.7\times 10^{-6}$. Meanwhile, the lightest neutrino mass and three CP-violating phases are predicted to be $m^{}_1 = 5.73$ meV, $\delta^{}_{\rm CP} = 242.99^\circ$, $\rho = 8.58^\circ$ and $\sigma = 92.27^\circ$, respectively.
  \item As can seen from Table~\ref{tab:parameter-space} and also Fig.~\ref{fig:para}, there are two disconnected ranges for $m^{}_1$, $\delta^{}_{\rm CP}$, $\rho$ and $\sigma$. The two disconnected ranges for $m^{}_1$, $\rho$ and $\sigma$ can be easily understood in the limit of the $\mu$-$\tau$ reflection symmetry (i.e., $\varepsilon = 0$), where two possible values of $m^{}_1$ respectively associated with two specific combinational values of $\rho$ and $\sigma$ can be worked out by Eq.~(\ref{eq:relation-2}), that is $m^{}_1=2.2$ meV for $(\rho, \sigma)=(\pi/2, 0)$ and $m^{}_1 = 6.3$ meV for $(\rho,\sigma) = (0, \pi/2)$, as discussed above. Thus when the $\mu$-$\tau$ reflection symmetry is slightly broken, (in other words, $\varepsilon$ is small but not zero), the values of $m^{}_1$, $\rho$ and $\sigma$ will be around these two possible solutions in the limit of $\mu$-$\tau$ reflection symmetry, which is exactly the case considered here. One may understand the two disconnected ranges for $\delta^{}_{\rm CP}$ with the help of Eq.~(\ref{eq:deviation-approx}). It can be easily checked that when the best-fit values of $\Delta m^2_{21}$ and $\theta^{}_{12}$ in Eq.~(\ref{eq:ex-neu-mixing}) are taken into consideration, $1-r^{}_{12} -(1+r^{}_{12}) \cos2\theta^{}_{12} $ is monotonically decreasing as a function of $m^{}_1$ and $m^{}_1 \simeq 4$ meV is a zero of it. This leads to $\delta^{}_{\rm CP}>270^\circ$ for $m^{}_1 \lesssim 4$ meV and $\delta^{}_{\rm CP} < 270^\circ$ for $m^{}_1 \gtrsim 4$ meV with initially $\delta^{}_{\rm CP} = 270^\circ$ in the limit of $\mu$-$\tau$ reflection symmetry, which is consistent with the obtained numerical results for $\delta^{}_{\rm CP}$.
  \item The upper-left panel of Fig.~\ref{fig:para} shows that the value of $\Delta \theta^{}_{23}$ is approximately proportional to $-\varepsilon$, and $\theta^{}_{23}$ is in the second octant for $\varepsilon < 0$. These behaviors of $\theta^{}_{23}$ coincide with the approximate analytical formula for the deviation of $\theta^{}_{23}$ given in Eq.~(\ref{eq:deviation-approx}), but in order to explain the two parameter spaces for $\theta^{}_{23}$-$\varepsilon$, higher order terms involving $r^{}_{12}$ in the analytical result for $\Delta \theta^{}_{23}$ need to be restored. The deviation of $\delta^{}_{\rm CP}$ shown in the upper-middle panel of Fig.~\ref{fig:para} is much larger than other mixing parameters, which is actually enhanced by $1/\theta^{}_{13}$ as revealed in Eq.~(\ref{eq:deviation-approx}). From the lower-right panel of Fig.~\ref{fig:para}, one can find that the maximal value $r$ can take for $m^{}_1 \sim 2.5$ meV is smaller than that for $m^{}_1 \sim 5.5$ meV. This may be roughly understood with the help of Eq.~(\ref{eq:r-approx}).
  \item The approximate result for $\Delta a^{}_\mu$ in Eq.~(\ref{eq:dmm}) where $\Delta a^{}_\mu$ is a quadratic function of the model parameter $a$, perfectly describes the behavior of $\Delta a^{}_\mu$ against $a$ shown in the left panel of Fig.~\ref{fig:para1}. However, the approximate formula for the branching ratio $\mathcal{B} \left( \tau^- \to \mu^- + \gamma \right)$ given in Eq.~(\ref{eq:bra-approx}) predicts smaller values than those shown in the right panel of Fig.~\ref{fig:para1}. The reason is that for $r\sim 10^{-7}$, the masses of the physical $S^{}_1$ and $\tilde{R}^{-\frac{1}{3}}_2$ are highly degenerate, leading to severe cancellation between the chirally enhanced contributions to $\mathcal{B} \left( \tau^- \to \mu^- + \gamma \right)$ from the physical $S^{}_1$ and $\tilde{R}^{-\frac{1}{3}}_2$ with $x$ given in Eq.~(\ref{eq:x}), and then the contributions not enhanced (e.g., the second terms in the square brackets in Eq.~(\ref{eq:ampSALR})) dominate the value of $\mathcal{B} \left( \tau^- \to \mu^- + \gamma \right)$ in the present case.
\end{itemize}

\section{Summary}\label{sec:conc}
For the purpose of giving a combined explanation of the tiny neutrino masses, lepton flavor mixing and muon $g-2$, we have extended the SM with two TeV-scale scalar leptoquarks $S^{}_1$ and $\widetilde{R}^{}_2$. After constructing the complete Lagrangian with baryon number conservation, we calculate the neutrino mass matrix generated via one-loop quantum corrections, where the mixing between $S^{}_1$ and $\widetilde{R}^{}_2$ resulting from the lepton-number-violating LQ-Higgs interaction plays a greatly significant role. We also recalculate all the contributions to the radiative decays of charged leptons and the muon $g-2$ from all leptoquarks. The contributions from the physical $S^{}_1$ and $\widetilde{R}^{-\frac{1}{3}}_2$ are quite similar and both largely enhanced by the up-type quark masses, which dominate the branching ratios of charged-lepton radiative decays and the size of the muon $g-2$. Given the textures of the leptoquark Yukawa coupling matrices shown in Eq.~(\ref{eq:Yukawa-matrices}), the neutrino mass matrix possesses an approximate $\mu$-$\tau$ reflection symmetry, and moreover its $(1,1)$ element vanishes, i.e., $\left( M^{}_\nu \right)^{}_{ee} = 0$, which is only in favor of the normal neutrino mass ordering. Considering this novel texture of the neutrino mass matrix, we have analytically discussed its main features and the associated neutrino masses and lepton flavor mixing in detail. Finally, the numerical calculations show that our model can successfully explain the anomaly of the muon $g-2$ and accommodate the latest global-fit results of neutrino oscillation data, as well as satisfy the constraints from the radiative decays of charged leptons. More specifically, given the best-fit values of the model parameters with $\chi^2_{\rm min} = 6.7 \times 10^{-6}$, the central value of $\Delta a^{}_\mu$ and the best-fit values of $\Delta m^2_{21}$, $\Delta m^2_{31}$, $\theta^{}_{12}$, $\theta^{}_{13}$ and $\theta^{}_{23}$ respectively given in Eqs.~(\ref{eq:ex-dam}) and (\ref{eq:ex-neu-mixing}) are reproduced pretty well, and additionally the lightest neutrino mass and three CP-violating phases are predicted to be $m^{}_1 = 5.73$ meV, $\delta^{}_{\rm CP} = 242.99^\circ$, $\rho = 8.58^\circ$ and $\sigma = 92.27^\circ$. An interesting feature of our model is that the allowed ranges of $m^{}_1$, $\rho$ and $\sigma$ are highly correlated, i.e., $m^{}_1 \sim 2.2$ meV for $(\rho,\sigma) \sim (\pi/2,0)$ and $m^{}_1 \sim 6.3$ meV for $(\rho, \sigma) \sim (0, \pi/2)$, as a result of the approximate $\mu$-$\tau$ reflection symmetry and one-zero texture of $M^{}_\nu$.

It is worth remarking that leptoquarks may naturally originate from the grand unification framework, and the leptoquark extensions of the SM can not only give a combined explanation of neutrino masses, lepton flavor mixing and the muon anomalous magnetic moment, as focused on in this work, but also have the potential to address flavor anomalies in the quark sector. Therefore, as appealing and promising extensions of the SM, leptoquark models deserve more attentions and further investigations.

\section*{Acknowledgements}

I greatly appreciate the encouragements and useful suggestions from Prof. Zhi-zhong Xing, and also thank Yu-feng Li, Xin Wang and Shun Zhou for helpful discussions. This research work is partly supported by the National Natural Science Foundation of China under grant No. 11775231, No. 11835013 and No. 12075254.

\end{document}